%% file: main.tex
\documentclass[
    superscriptaddress,
    floatfix,
    reprint,
    aip,
    jcp,
]{revtex4-2}

\input{preamble}
\myexternaldocument[si-]{si}

\begin{document}

\title{\manuscripttitle}

\input{authors}



\input{main/sections/abstract}

\maketitle

\input{main/sections/introduction}
\input{main/sections/background}
\input{main/sections/method}

\input{main/sections/application}
\input{main/sections/conclusion}
\input{main/sections/supplementary}
\input{main/sections/acknowledgments}
\input{main/sections/data_availability}

\bibliography{bibliography/kevin, bibliography/philip, bibliography/egor, bibliography/michele, bibliography/qianjun}

\end{document}


\title{\manuscripttitle:~Supplementary~Material}

\input{authors}

\maketitle
\tableofcontents

\nomenclature{}{}

\clearpage\input{si/sections/nomenclature}
\clearpage\input{si/sections/conventions}
\clearpage\input{si/sections/ewald_derivation}
\clearpage\input{si/sections/lode}
\clearpage\input{si/sections/accuracy}
\clearpage\input{si/sections/cutoff_search}
\clearpage\input{si/sections/gpu-utilization}
\clearpage\input{si/sections/water-nve}

%

%% file: preamble.tex
\usepackage{amsthm}
\usepackage{amssymb}
\usepackage[english]{babel}
\usepackage{color}
\usepackage{comment}
\usepackage{hyperref}
\usepackage[utf8]{inputenc}
\usepackage[fleqn]{mathtools}
\usepackage{sidecap}
\usepackage{svg}
\usepackage{capt-of}
\usepackage{xr}
\usepackage{xfrac}
\usepackage{xspace}
\usepackage{soul} 

\newcommand{\manuscripttitle}{Fast and flexible long-range models for atomistic machine learning}

\theoremstyle{definition}

\theoremstyle{remark}

\newcommand{\mybold}[1]{\mathbf{#1}}

\newcommand{\bk}{\mybold{k}}
\newcommand{\br}{\mybold{r}}

\newcommand{\rcut}{r_\mathrm{cut}}

\newcommand{\torchpme}{\emph{torch-pme}\xspace}
\newcommand{\jaxpme}{\emph{jax-pme}\xspace}
\newcommand{\torch}{\emph{PyTorch}\xspace}
\newcommand{\jax}{\emph{JAX}\xspace}
\newcommand{\forceunits}{\mathrm{kcal/mol/\AA}}

\newcommand{\primesum}{\sideset{}{'}\sum}

\newcommand{\rev}[1]{#1}
\makeatletter
\newcommand*{\addFileDependency}[1]{
  \typeout{(#1)}
  \@addtofilelist{#1}
  \IfFileExists{#1}{}{\typeout{No file #1.}}
}\makeatother

\newcommand*{\myexternaldocument}[2][]{%
\externaldocument[#1]{#2}%
\addFileDependency{#2.tex}%
\addFileDependency{#2.aux}%
}

%% file: authors.tex
\newcommand{\cosmo}{\affiliation{Laboratory of Computational Science and Modeling, IMX, École Polytechnique Fédérale de Lausanne, 1015 Lausanne, Switzerland}}
\newcommand{\marvel}{\affiliation{National Centre for Computational Design and Discovery of Novel Materials (MARVEL), {\'E}cole Polytechnique F{\'e}d{\'e}rale de Lausanne, 1015 Lausanne, Switzerland}
}

\author{Philip~Loche}
\cosmo
\marvel
\thanks{These two authors contributed equally.}
\author{Kevin~K.~Huguenin-Dumittan}
\cosmo
\thanks{These two authors contributed equally.}
\author{Melika~Honarmand}
\cosmo
\marvel
\author{Qianjun~Xu}
\cosmo
\author{Egor~Rumiantsev}
\cosmo
\marvel
\author{Wei~Bin~How}
\cosmo
\marvel
\author{Marcel~F.~Langer}
\cosmo
\author{Michele~Ceriotti}
\cosmo
\marvel
\email{michele.ceriotti@epfl.ch}

%% file: main/sections/abstract.tex
\begin{abstract}
  Most atomistic machine learning (ML) models rely on a locality ansatz, and decompose
  the energy into a sum of short-ranged, atom-centered contributions. This leads to
  clear limitations when trying to describe problems that are dominated by long-range
  physical effects -- most notably electrostatics. Many approaches have been proposed to
  overcome these limitations, but efforts to make them efficient and widely available
  are hampered by the need to incorporate an ad hoc implementation of methods to treat
  long-range interactions. We develop a framework aiming to bring some of the
  established algorithms to evaluate non-bonded interactions  -- including Ewald
  summation, classical particle-mesh Ewald (PME), and particle-particle/particle-mesh
  (P3M) Ewald -- into atomistic ML. We provide a reference implementation for \torch as
  well as an experimental one for \jax. Beyond Coulomb and more general long-range
  potentials, we introduce purified descriptors which disregard the immediate
  neighborhood of each atom, and are more suitable for general long-ranged ML
  applications. Our implementations are fast, feature-rich, and modular: They provide an
  accurate evaluation of physical long-range forces that can be used in the construction
  of (semi)empirical baseline potentials; they exploit the availability of automatic
  differentiation to seamlessly combine long-range models with conventional, local ML
  schemes; and they are sufficiently flexible to implement more complex architectures
  that use physical interactions as building blocks. We benchmark and demonstrate our
  \torchpme and \jaxpme libraries to perform molecular dynamics simulations, to train
  \rev{multiscale} ML potentials, and to evaluate long-range equivariant descriptors of
  atomic structures.
\end{abstract}

%% file: main/sections/introduction.tex
\section{Introduction}

Machine learning (ML) has become essential in atomistic simulations, especially as a
surrogate for quantum mechanical methods like density functional theory (DFT). Early
works by Behler and Parrinello~\cite{Behler2007} and  Bartók et al.~\cite{Csanyi2010}
demonstrated that ML models are able to predict the energy and forces of an atomic
structure at close to DFT accuracy with significantly reduced computational cost.
Since then, the field has rapidly expanded in many directions: On one hand, various
architectures including kernel-based approaches~\cite{Csanyi2010, Rupp2012, Schutt2018}
and message-passing neural networks have emerged to further improve accuracy and
efficiency~\cite{SchNet, DimeNet, PaiNN, Thomas2018, musaelian_learning_2023}.
Furthermore, beyond just the energy or other rotationally invariant target properties,
the field has seen the rise of equivariant model architectures that can also handle
tensorial target properties such as dipoles and polarizabilities ~\cite{glie+17prb, grisafi_symmetry-adapted_2018, wilkins_accurate_2019},
the electron
density~\cite{broc+17nc,fabrizio_electron_2019, lewis_learning_2021}, the Hamiltonian
operator~\cite{schu+19nc,cignoni_2024} and many-particle wavefunctions~\cite{carl-troy17science,herm+20nc,unke_se3-equivariant_2021}. Thanks to
the mathematical field of representation theory, there is now a systematic understanding
about how to incorporate any type of behavior under rotations in ML models~\cite{fulton1991representation, serre1977linear, tinkham2003group}.

Despite these advancements, a persistent limitation in many of the ML models is the use
of finite cutoff radii, which had been justified in terms of the ``nearsightedness of
electronic matter''~\cite{kohn_density_1996, prodan_nearsightedness_2005}. While this
allows for efficient methods with computational costs that scale linearly with the
number of atoms, the trade-off is the introduction of systematic errors in scenarios
where long-range interactions play a crucial role. For instance, electrostatic
interactions from the direct Coulomb potential to induced multipolar interactions are critical in ionic and polar materials, affecting dielectric constants,
phonon spectra, and structural stability~\cite{Fennell2006, Born1921, jackson_classical_1998}. Dispersion (van
der Waals) forces are equally vital in layered materials, molecular crystals and
interfaces, as they contribute to cohesion and adhesion energies~\cite{Grimme2006,
Tkatchenko2009, DiStasio2014}.
Ignoring these interactions
in ML models restricts their applicability.

This work presents libraries for the \torch and \jax frameworks that implement efficient methods for computing a family of long-range features as well as classical long-range energy corrections in an automatically differentiable manner. Due to the use of popular and flexible ML libraries, the features can easily be integrated into any pre-existing ML framework for invariant and equivariant target properties, with or without gradients.

In \autoref{sec:background}, we begin by providing an overview of the scientific background concerning algorithms for long-range electrostatics developed by the molecular dynamics (MD) community, as well as long-ranged ML methods.
In \autoref{sec_method}, we discuss some of the specific technical choices and the core features of our implementation.
In  \autoref{sec:results}, we demonstrate and benchmark our reference \torchpme implementation in several different use cases, providing also some comparisons with its more experimental \jaxpme companion.

%% file: main/sections/background.tex
\section{Background}\label{sec:background}

Long-range interactions are not new to ML, and have in fact already been studied
extensively since the outset of traditional atomistic simulations. Here we briefly summarize the key ideas.

\subsection{Notation and Conventions}

In this subsection, we focus on the Coulomb potential, and define the bare potential function
\begin{align}
    v(r) = \frac{1}{r}
\end{align}
for $r\in (0,\infty)$, even though the general setting of the problem applies to any pair potential, and we will discuss a few concrete cases of other $1/r^p$-type interactions for more general exponents $p>0$.
In a system of $N$ atoms with positions $\br_i \in \mathbb{R}^3$ and weights (charges) $q_i$ for $i=1,\dots,N$, the Coulomb potential $V_i$ at the location of an atom $i$ and the total energy $E$ of the system can be defined as
\begin{align}
  V_i & =  \frac{1}{2}\sum_{j\neq i} q_j v(r_{ij}) \\
  E & = \sum_{i=1}^N q_i V_i \,,
\end{align}
where $r_{ij} = \|\br_i - \br_j\|$ is the distance between atoms $i$ and $j$, and we also include the factor of $1/2$ to account for double-counting directly in the definition of $V_i$.

We can extend the atomic potentials to periodic systems by also including all interactions with periodic images of the atoms. Thus, 
\begin{align}
    V_i = \frac{1}{2}\sum_{j=1}^N \primesum_{\mybold{l}} q_j v \left(\|\br_i - \br_j - \mybold{l}\|\right)\,,
w\end{align}
where $\mybold{l}$ runs over the lattice vectors of the periodic cell. The sum over $j$ also includes $i=j$ in the periodic case to take into account the interactions between atom $i$ and its own periodic images. For this special case of $i=j$, the prime symbol in the second summation sign indicates that the term with $\mybold{l}=0$ must be omitted \rev{to avoid the singularity in $v$ as $r\rightarrow 0$.}

While conceptually straight-forward, this definition of the potential in periodic space is problematic for the Coulomb potential since its slow decay makes the sum over $\mybold{l}$ conditionally convergent. A naive implementation (such as spherical summation) does not even converge~\cite{Borwein_Glasser_McPhedran_Wan_Zucker_2013}.

\subsection{Ewald Summation}
Ewald summation, published in 1921, remains a foundational method that effectively defines how
to extend this computation to periodic systems~\cite{Ewald1921}. The proposed solution is to perform a range separation of the Coulomb potential: We write
\begin{align}
    v(r) & = v_\mathrm{SR}(r) + v_\mathrm{LR}(r) \\
    v_\mathrm{SR}(r) & = \frac{\mathrm{erfc}\left(\frac{r}{\sqrt{2}\sigma}\right)}{r} \\
    v_\mathrm{LR}(r) & = \frac{\mathrm{erf}\left(\frac{r}{\sqrt{2}\sigma}\right)}{r}\,,
\end{align}
where $\sigma$ is a ``smearing'' parameter that defines the range separation and $\mathrm{erf}$ ($\mathrm{erfc}$) is the (complementary) error function $\mathrm{erf}(x) = 2/\sqrt{\pi}\int_0^x \mathrm{exp}\left(-t^2\right)\,\mathrm{d}t$ (and $\mathrm{erfc}[x] = 1 - \mathrm{erf}[x]$). The first term $v_\mathrm{SR}$ contains the singularity of the Coulomb potential as $r\rightarrow 0$ and decays quickly as $r\rightarrow\infty$. Thus, we call this the short-range part of the Coulomb potential. $ v_\mathrm{LR}$ on the other hand decays as $1/r$ as $r\rightarrow \infty$ just like the Coulomb potential, but remains smooth and finite for $r\rightarrow 0$. We call it the long-range part of the Coulomb potential. The transition between the two happens roughly at $r\approx\sigma$. We will shortly discuss how to choose $\sigma$ in practice.

Using this range separation, the potential of atom $i$ can be written as
\begin{align}
    V_i & = \frac{1}{2}\sum_{j=1}^N \primesum_{\mybold{l}}q_jv_\mathrm{SR}\left(\|\br_i - \br_j - \mybold{l}\|\right) \\
    & + \frac{1}{2}\sum_{j=1}^N \primesum_{\mybold{l}}q_jv_\mathrm{LR}\left(\|\br_i - \br_j - \mybold{l}\|\right)\,.
\end{align}
Due to the fast decay of $v_\mathrm{SR}$, the sum over $j$ and $\mybold{l}$ in the first term converges quickly. We can thus efficiently implement it by introducing a cutoff radius $r_\mathrm{cut}$, and only summing up the contributions up to this distance.

The second sum is still conditionally convergent, since $v_\mathrm{LR}$ contains the slowly decaying behavior of the Coulomb potential. However, since the singularity as $r\rightarrow 0$ has been removed, $v_\mathrm{LR}$ is now a smooth function and hence has a quickly decaying Fourier transform given by
\begin{align}\label{eq:vhat_lr_coulomb}
    \hat{v}_\mathrm{LR}(k) = \frac{4\pi}{k^2}e^{-\frac{1}{2}\sigma^2k^2} \,;
\end{align}
$k = \|\bk\|$ is the norm of the reciprocal space vector $\bk$.

We can now use the Poisson summation formula, which allows us to convert a sum over a periodic lattice into a dual sum in reciprocal space. The details on the used conventions for Fourier transforms and a proof of the Poisson summation formula for the convenience of readers can be found in section S1 of the supporting information. The final result for the potential of atom $i$ is
\begin{align}
    V_i & = V_{i,\mathrm{SR}} + V_{i,\mathrm{LR}} +  V_{i,\mathrm{self}} + V_{i,\mathrm{charge}} \\
    V_{i,\mathrm{SR}} & = \frac{1}{2}\sum_{j=1}^N \primesum_{\mybold{l}} q_jv_\mathrm{SR}\left(\|\br_i - \br_j - \mybold{l}\|\right) \\
    \label{potential_lr} V_{i,\mathrm{LR}} & = \frac{1}{2\Omega}\sum_{j=1}^N \sum_{\mybold{k}\neq 0}q_j \hat{v}_\mathrm{LR}(k) e^{i\mybold{k}(\br_i-\br_j)} \\
    V_{i,\mathrm{self}} & = -\frac{1}{2} q_i \sqrt{\frac{2}{\pi}}\frac{1}{\sigma} \\
    V_{i,\mathrm{charge}} & = -\frac{\pi\sigma^2 Q}{\Omega}\,, 
\end{align}
where $\Omega$ is the volume of the unit cell, the sum over $\mybold{k}$ is a sum over all nonzero reciprocal space lattice vectors and $Q = \sum_i q_i$ is the total charge of all atoms in the unit cell. We have kept the overall factor of $1/2$ that arises from double counting at the front for better comparison with results from some authors which \rev{handle this factor differently.} The extra term $V_{i,\mathrm{charge}}$ is required to enforce charge neutrality, a requirement of the method. If the provided charges $q_i$ do not lead to a charge-neutral cell, a homogeneous background charge is added to the system to ensure overall neutrality in the unit cell. The term is then the interaction between an atom and this background charge. $V_{i,\mathrm{self}} = q_iv_\mathrm{LR}(0)/2$ is the self-term which compensates the interaction of an atom with itself via $v_\mathrm{LR}$, which is an artifact of the reciprocal space sum.

Due to the fast decay of $\hat{v}_\mathrm{LR}(k)$ as $k\rightarrow\infty$, the long-range sum can now also be computed efficiently by choosing a cutoff $k_\mathrm{cut}$ and truncating the sum by only including terms for which $k \leq k_\mathrm{cut}$. The naive Ewald algorithm has a computational cost scaling quadratically with the number of atoms, $\mathcal{O}\left(N^2\right)$,
while optimizations -- letting $r_\mathrm{cut}$ scale with the size of the unit cell -- can bring the exponent in the scaling down to
$\mathcal{O}\left(N^{\sfrac{3}{2}}\right)$~\cite{allentildesley}.

\subsection{Generalization to Arbitrary Exponents}

\rev{This} work has been extended a few decades later to other
interactions \rev{beyond the Coulomb potential}, including more general power-law potentials $v(r)=1/r^p$ with an
interaction exponent $p>0$~\cite{nijboer_calculation_1957, williams_accelerated_1971, williams_accelerated_1989, karasawa1989acceleration}. The Coulomb interaction corresponds to $p=1$, while dispersion corresponds to $p=6$. Summarizing their results using our
conventions and notations, the range separation $v = v_\mathrm{SR} + v_\mathrm{LR}$ is
performed using
\begin{align}
    v_\mathrm{SR}(r) & = \frac{1}{\Gamma\left(\frac{p}{2}\right)}\frac{\Gamma\left(\frac{p}{2},\frac{r^2}{2\sigma^2}\right)}{r^p} \\
    v_\mathrm{LR}(r) & = \frac{1}{\Gamma\left(\frac{p}{2}\right)}\frac{\gamma\left(\frac{p}{2},\frac{r^2}{2\sigma^2}\right)}{r^p} \\
    \label{eq:vhat_lr_general} \hat{v}_\mathrm{LR}(k) & = \frac{\pi^{3/2}2^{3-p}}{\Gamma\left(\frac{p}{2}\right)}\frac{\Gamma\left(\frac{3-p}{2}, \frac{1}{2}\sigma^2k^2\right)}{k^{3-p}} \,,
\end{align}
where $\Gamma(a)=\int_0^\infty t^{a-1}e^{-t}\,\mathrm{d}t$ is the (complete) Gamma function, $\Gamma(a,x)=\int_x^\infty t^{a-1}e^{-t}\,\mathrm{d}t$ is the upper incomplete Gamma function and $\gamma(a,x) = \Gamma(a) - \Gamma(a,x) = \int_0^x t^{a-1}e^{-t}\,\mathrm{d}t$ the lower incomplete Gamma function.
Up to a prefactor, we see that the functional form of $\hat{v}_\mathrm{LR}$ is identical to $v_\mathrm{SR}$, the difference being that $r/\sigma$ gets replaced by $\sigma k$ and $p$ by $3-p$. This high degree of symmetry is one reason that makes this particular choice of $v_\mathrm{SR}$ efficient. \rev{Note that we use a custom  implementation of the incomplete Gamma functions that, contrary to the one present in \torch, supports integer exponents up to $p=6$.  }

To get the atomic potentials, the decomposition
\begin{align}
    V_i & = V_{i,\mathrm{SR}} + V_{i,\mathrm{LR}} + V_{i,\mathrm{self}} + V_{i,\mathrm{charge}}
\end{align}
is still valid, where the first two terms are computed using the identical method but using the generalized expressions for $v_\mathrm{SR}$ and $\hat{v}_\mathrm{LR}$, while
\begin{align}
    V_{i,\mathrm{self}} & = -\frac{q_i}{2}\frac{1}{(2\sigma^2)^\frac{p}{2}\Gamma(\frac{p}{2}+1)} \\
    V_{i,\mathrm{charge}} & = -\frac{\pi^{\frac{3}{2}}(2\sigma^2)^{\frac{3-p}{2}}}{(3-p)\Gamma(\sfrac{p}{2})}\frac{Q}{\Omega} \, .
\end{align}

\rev{We remark that the self-term is always necessary for any exponent $p\in(0,\infty)$. The charge correction term, on the other hand, should only be used for $p<3$. This is because, only for these slowly decaying potentials, the energy diverges if the cell is not ``charge-neutral'', thus requiring the addition of a homogeneous background, while for $p>3$ this is not the case. The critical case $p=3$ is more subtle, but we treat it just like $p>3$, i.e. by setting $V_{i,\mathrm{charge}}=0$. A derivation of these two correction terms with a more detailed discussion of the their subtleties can be found in the supporting information in section S2.
Note that neither of the terms depends on the particle positions, and hence do not affect gradients or forces.}

\subsection{Mesh-Based Approaches}
The Ewald sum provided the first absolutely convergent formulation for the potential in a periodic system, and remains important to this day. It suffers, however, from a relatively inefficient computational scaling with the number of atoms. Thus, many algorithms have been developed to significantly reduce the computational
cost of the Ewald sum, enabling the use of long-range methods for the simulation of
large-scale systems. The most important family for this work consists of the
particle-mesh methods, such as particle-particle
particle-mesh (P3M) which was published first \cite{Hockney1981, deserno_how_1998}, the original particle mesh Ewald (PME) \cite{Darden1993} and its extension called smooth PME (SPME)  \cite{essmann1995smooth}. \rev{These methods use the fast Fourier transform (FFT)\cite{winkler_numerical_1993}} to perform the reciprocal space part of the Ewald sum, and bring the asymptotic scaling of the computational cost down to
$\mathcal{O}(N\log N)$. They assume
periodicity in the system, but can also be used for aperiodic systems with some
care and modifications~\cite{kawata2001particle}.

The key idea is that the two sums appearing in \autoref{potential_lr} can ``almost'' be interpreted as discrete Fourier transforms: We can in fact break down the computation of $V_{i,\mathrm{LR}}$ into the two sums
\begin{align}
    \label{eq:fft1} \hat{q}(\bk) & = \sum_{j=1}^N q_j e^{-i\bk\br_j} \\
    \label{eq:fft2} V_{i,\mathrm{LR}} & = \sum_{\mybold{k}\neq 0}\hat{q}(\bk)\hat{v}_\mathrm{LR}(k)e^{i\bk\br_i} \,.
\end{align}

Both \autoref{eq:fft1} and \autoref{eq:fft2} look quite similar to Fourier transforms, but with one key difference: The positions $\br_i$ and $\br_j$ are in general not equally spaced. Thus, despite the apparent similarity, these cannot directly be computed using algorithms for the discrete Fourier transform.

The similarity however motivates the following trick: In order to use the efficient FFT algorithm, we can interpolate the weights $q_i$ onto an equally-spaced mesh. This generates an effective (charge) density $\rho$ that is only defined on the mesh. While this changes the position of the particles, and hence introduces discretization errors, the regular nature of the particle positions now allows us to compute the appropriately modified versions of both \autoref{eq:fft1} and \autoref{eq:fft2} efficiently using three-dimensional FFT. The computational cost typically scales as $\mathcal{O}(N\log N)$, assuming a fixed mesh spacing and that the system volume is proportional to the number of atoms. This general description and scaling is common to all mesh-based methods.

The PME, P3M and SPME methods simply differ in (1) the choice of interpolation scheme between the particle positions and the mesh, (2) the choice of $\hat{v}_\mathrm{LR}$ that is used in \autoref{eq:fft2}, and (3) the method to perform the differentiation to compute the electric field or forces. We briefly summarize the first two key distinguishing points for PME and P3M, as these are particularly relevant for this work. Deserno and Holm \cite{deserno_how_1998} provide a more detailed discussion.

The original version of PME \cite{Darden1993} is the simplest for both (1) and (2). The interpolation is performed using (piecewise) Lagrange interpolation, a well-known technique in numerical analysis \cite{gautschi2011numerical}. Furthermore, $\hat{v}_\mathrm{LR}$ remains the same as for the Ewald summation, \autoref{eq:vhat_lr_coulomb} for Coulomb and \autoref{eq:vhat_lr_general} for the general case. It can in fact be shown that for this choice of $\hat{v}_\mathrm{LR}$, the Lagrange interpolation scheme is optimal regarding the discretization errors arising due to the mesh \cite{deserno_how_1998}.

This is the simplest of the three variants to implement, and therefore serves as a convenient testing ground for new implementations. One key downside, however, is that the interpolation scheme leads to non-smooth behavior or even discontinuities \rev{even for arbitrarily high interpolation orders \cite{deserno_how_1998}}, making the method less suitable for our approach that fully exploits the \rev{automatic differentiation capabilities} of \torch and \jax. The authors of PME have in fact addressed this issue by presenting the SPME algorithm just two years after the original version \cite{essmann1995smooth}.

The P3M algorithm, despite being more complex than the original PME, was historically presented first. It is widely used due to its superior accuracy \cite{deserno_how_1998}. The key idea is that in a periodic system, the effective Coulomb potential \rev{or Green's function} is different than in vacuum due to the periodic neighbors. This is what suggests a modification of $\hat{v}_\mathrm{LR}$ by taking into account the periodic cell. Since the crystal lattice breaks rotation symmetry, $\hat{v}_\mathrm{LR}(\bk)$ no longer is spherically symmetric and now depends on the three-dimensional vector $\bk$ rather than just its norm $k$. In practice, $\hat{v}_\mathrm{LR}$ is obtained by minimizing an estimated force error due to discretization. P3M further uses a custom interpolation scheme that is obtained from requiring a smooth interpolation. The combination of the interpolation scheme and optimized $\hat{v}_\mathrm{LR}$ makes P3M more accurate, and hence a popular choice despite its higher complexity compared to naive PME.

Beyond the mesh-based ones, many other algorithms have been developed to efficiently compute electrostatic interactions, including the fast
multipole method that was the first linear-scaling, $\mathcal{O}(N)$ method for
electrostatics~\cite{Greengard1987, Gibbon1992}. Despite the better asymptotic
scaling \rev{and hence faster speed as $N\rightarrow\infty$}, the larger prefactor typically makes the particle-mesh methods more suitable
for the large system sizes \rev{of $N=10^3 \sim 10^5$}, which is why, in this work, we focus on particle-mesh methods rather than fast multipole methods.

\subsection{Long-Range ML Models}
Inspired by these developments for long-range interactions in the last century, it has become increasingly popular
for ML models in atomistic modeling to incorporate long-range corrections as well. In fact, one
of the founding works by Bartók et al.~\cite{Csanyi2010} already included an explicit
dispersion term in the prediction of the energy for GaAs structures. A similar strategy
has been used by many subsequent works \cite{deng_electrostatic_2019,deringer_general-purpose_2020,unke2021spookynet,niblett_learning_2021,kabylda2024molecular}, in which the energy and optionally the forces of
a structure are predicted as the sum of a more sophisticated ML part and an explicit
long-range term of the form $1/r^p$. Typically, each chemical species is assigned a fixed weight (or charge in the case of Coulomb interactions) in these interactions. This makes it simple
to train the models, and has the advantage that it can be combined with essentially any
model architecture.

In the last five years, the field has also seen the rise of more sophisticated
approaches going beyond simple point charge models to handle more complex datasets. One
extension is to make the particle weights or charges themselves dependent on the atomic
environments~\cite{Physnet2019, Sifain2018}, and to learn this relationship directly
from the data. Alternatively, it has been proposed to learn the positions of the Wannier centers
\cite{gao-rems22nc,peng_efficient_2023,zhang_deep_2022}. This allows the center of the electron
charge to be different from the positively charged ion, naturally capturing dipolar
effects. Alternatively, it is possible to learn electronegativities rather than the
charges, and to compute the charges from a subsequent charge equilibration scheme
\cite{faraji_high_2017, ko_fourth-generation_2021, ko_general-purpose_2021,
gubler_accelerating_2024}. This approach has the advantage that the final charges can
depend on atoms beyond a local cutoff, and hence can capture nonlocal charge transfer.
A general property of these methods is that they are most naturally designed to have the
energy (and forces) as the target property. This specialization makes the methods very
powerful for this task, but less suitable to other target properties. In addition, the
original implementation lacked efficiency and scalability, but this was recently
resolved by also using a mesh paradigm~\cite{gubler_accelerating_2024} improving the
scaling significantly.

\rev{While these approaches work well to incorporate electrostatic energy corrections, there are two common deficits.
First, being explicitly designed to reproduce physical interactions, they are not necessarily well-suited for the data-driven modeling of long-range effects with a different origin, or affecting a property other than the energy.
Second, even though they are designed to incorporate long-range interactions, the numerical value of the quantities they compute is dominated by short-range contributions. This happens because even though the Coulomb potential (or more general power-laws) at the position of an atom depend on atoms that are far away, they are still primarily determined by neighbor atoms that are close by. This contamination by short-range terms can be compensated by the presence of a conventional short-range ML model, but goes against the notion of treating the two parts separately.
}

An alternative approach, developed by some of the authors in this work, emphasizes
constructing general equivariant features that encode long-range
information, rather than focusing directly on the energy as a target property
\cite{grisafi_incorporating_2019, grisafi_multi-scale_2021,huguenin-dumittan_physics-inspired_2023}. These
long-distance equivariant (LODE) features can be applied in various ML models to predict
a range of target properties, from scalar invariants to higher-order equivariants. While
prior studies have primarily explored the method's capabilities, the implementation used
an Ewald-like summation approach~\cite{huguenin-dumittan_physics-inspired_2023}, leading to an
$\mathcal{O}(N^2)$ scaling. This quadratic scaling limits the method's applicability to
larger systems.

As is the case for the usual Coulomb potential, LODE in its original formulation is not a range separated descriptor, but
contains all short-range (i.e., within cutoff) and long-range contributions of a system. \rev{To address this issue, we will introduce the exterior potential features (EPFs) in the next section, which is} a modification of the original formulation that only includes \rev{the contributions from exterior atoms}~\cite{huguenin-dumittan_physics-inspired_2023, chong_prediction_2024}. A similar idea has already been applied to LODE features in previous work~\cite{huguenin-dumittan_physics-inspired_2023}. In addition, recently there is
strong evidence~\cite{gubler_accelerating_2024, cheng_latent_2024, faller_density-based_2024} that using a full spherical
expansion of the potential is not necessary to describe long-range effects in machine learning models,
but so far there is no rigorous investigation of these observations. A mathematical motivation thereof is provided in section S3 of the supporting information, which also connects the approaches in this work with our previous work.

%% file: main/sections/method.tex
\section{Method}\label{sec_method}

Our primary goal is to design a library to compute long-range interactions that can be easily integrated with existing short-range architectures, essentially providing an easy-to-use framework to build
range separated models for atomistic machine learning.
To this end, our reference \torchpme library provides

\begin{enumerate}
  \item A modular implementation of range separated potentials \rev{working for arbitrary unit cells including triclinic ones.}
  \item Full integration with \torch, featuring differentiable and learnable
    parameters,
  \item Efficient particle-mesh-based computation with automatic hyperparameter tuning,
  \item Pure long-range descriptors, free of short-range contributions,
  \item Support for arbitrary invariant and equivariant features and ML architectures.
\end{enumerate}

We discuss these features in general terms, without focusing too much on the specific Application Programming Interface (API), because similar ideas can be readily implemented with other ML frameworks, as we are currently doing with an experimental \jaxpme library.

\begin{figure}[tbh]
\includegraphics{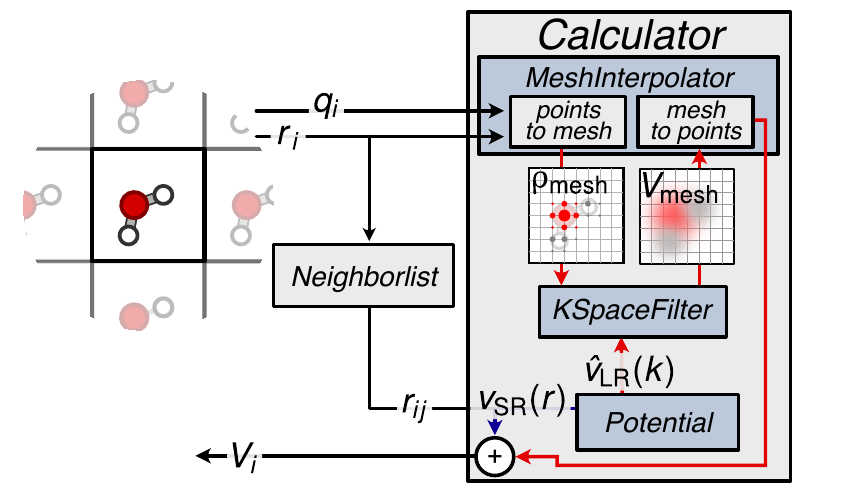}
  \caption{A schematic representation of the main building blocks 
    inside a \emph{Calculator} of a range separated architecture, that combines an
    evaluation of the short-range part of the potential $v_\mathrm{SR}(r)$ based on
    local interatomic distance information with the evaluation of the long-range part
    $v_\mathrm{LR}(k)$ using grids via a \emph{MeshInterpolator} and a
    \emph{KSpaceFilter}.
  }
  \label{fig:modular-scheme}
\end{figure}

\subsection{A modular implementation of range separated models}

As discussed in \autoref{sec:background}, the key concept underlying all efficient methods to model long-range interactions is to split the calculation of the pair interactions between two particles into a short and
long-range part, i.e., $v(r)=v_\mathrm{SR}(r)+v_\mathrm{LR}(r)$. The short-range part,
which decays to (effectively) zero beyond a set cutoff distance $r_{\text{cut}}$, is
computed by summing over the neighbors of each particle. The neighbor-list required for the pair distances $r_{ij}$ in the short-range part have to be calculated with an external library. These pairs are also needed for any short-range component of a ML model, and can be easily precomputed during training.
The long-range part is
computed in $k$-space, based on its Fourier transform $\hat{v}_\mathrm{LR}(k)$ and requires 
the absolute position vectors $\mybold r_i$ as well as the particle weights $q_i$ (that correspond to charges for an electrostatic potential).
Furthermore, the methods with an efficient scaling in the asymptotic limit, such as PME
or P3M, convert the particle-based description into a smooth particle density,
discretizing it on a real-space mesh, and using FFT to perform a Fourier convolution.

These ingredients can be used for more general tasks than computing explicitly physical
interatomic potentials. To make this possible, \rev{our implementation} follows a
modular design, in which a \emph{Potential} class exposes methods to compute $v(r)$,
$v_\mathrm{SR}(r)$, $v_\mathrm{LR}(r)$, $\hat{v}_\mathrm{LR}(k)$; a
\emph{MeshInterpolator} converts particle positions $\mybold{r}_i$ and pseudo-charges
$q_i$ into a density mesh, and vice-versa interpolates a scalar field defined
on the mesh at arbitrary positions; a \emph{KSpaceFilter} performs the Fourier-domain
convolution.

These elements can be combined in a \emph{Calculator} object that applies one of the
classical algorithms to compute long-range interactions, evaluating the potential at the
atomic positions, $V_i=V(\mybold{r}_i)$ (see \autoref{fig:modular-scheme}). 
However, the same modular
components can also be used in more complex ways.
For example, users can compute descriptors
based on the local potential in a neighborhood around each atom, such as LODE~\cite{grisafi_incorporating_2019,huguenin-dumittan_physics-inspired_2023} features (\autoref{sub:lode}).

As long as all components are implemented within an
auto-differentiable framework, both the base calculators and any custom architecture
allows to easily evaluate derivatives of the predictions with respect to particle
coordinates, cell matrix, and model parameters.

\subsection{Fully Differentiable and Learnable Framework}\label{flexibility_theory} 
In practice, the integration of long-range calculators with ML libraries built around automatic evaluation of gradients makes it easy to combine long-range terms with any \torch- and \jax-based ML model. 
This includes optimizing parameters such as the particle weights $q_i$ and even the type of
interaction itself by e.g. parametrizing the interaction as $v(r) = w_1/r
+w_2/r^2 + w_3/r^3$, where $w_1$, $w_2$, and $w_3$ are learnable. This flexibility is illustrated for both simple toy systems (\autoref{subsec_learnability}) and  complex model architectures  (\autoref{subsec_architectures}).
Automatic differentiation also makes the implementation of models that learn both the energies and forces effortless.

\subsection{Hyperparameter Tuning}\label{subsec_tuning_theory}

Range-separated calculations like Ewald \rev{summation} and its mesh-based variants require careful distribution
of the computational load between the real-space and the Fourier-space contributions. This is relevant both for computational performance
and correctness; poor choices of parameters can lead to severe errors of the potential and the forces, while over-emphasizing correctness can result in prolonged computational time. Parameters to be optimized include the mesh spacing and the real-space interaction cutoff. 
Our package comes with a parameter tuning functionality
 that simplifies the process for non-experts for the Coulomb potential
 based on known estimates for the force error $\mathcal{E}_F$~\cite{ewald_tuning_1995, pme_tuning_1995, desernoHowMeshEwald1998a}. 
 To this end, the relevant parameters, which we call $\mybold{\theta}$, are optimized to fulfill the target accuracy
 requirement, i.e. $\mathcal{E}_F(\mybold{\theta}) < \epsilon_\mathrm{target}$, while taking the shortest computational time.

 The accuracy can be set by the user, with some default
 suggestions. The cutoff $r_\mathrm{cut}$ is set to a fixed value by the user. In principle, the cutoff can also be optimized but its value is usually fixed by a short range machine learning model. The smearing $\sigma$ is optimized to achieve
 a real space error equal to $\epsilon_\mathrm{target} / 2$ using a closed-form estimate of the error. For the remaining parameters related to the $k$-space evaluation, a grid search is performed using a set of heuristically determined parameters. The force error
 estimate $\mathcal{E}$ also depends on the algorithm used, as its $k$-space part is different for Ewald
 summation, PME, and P3M. A mock computational speed test is performed for each set of parameters fulfilling the accuracy requirement.
 Further details and the concrete mathematical expressions can
 be found in section S4 of the supporting information. Benchmark results
 in \autoref{subsec_speed_and_accuracy} demonstrate and compare the obtained accuracy
 with the desired value $\epsilon_\mathrm{target}$, as well as the resulting speed.

\begin{figure}
  \includegraphics[width=\columnwidth]{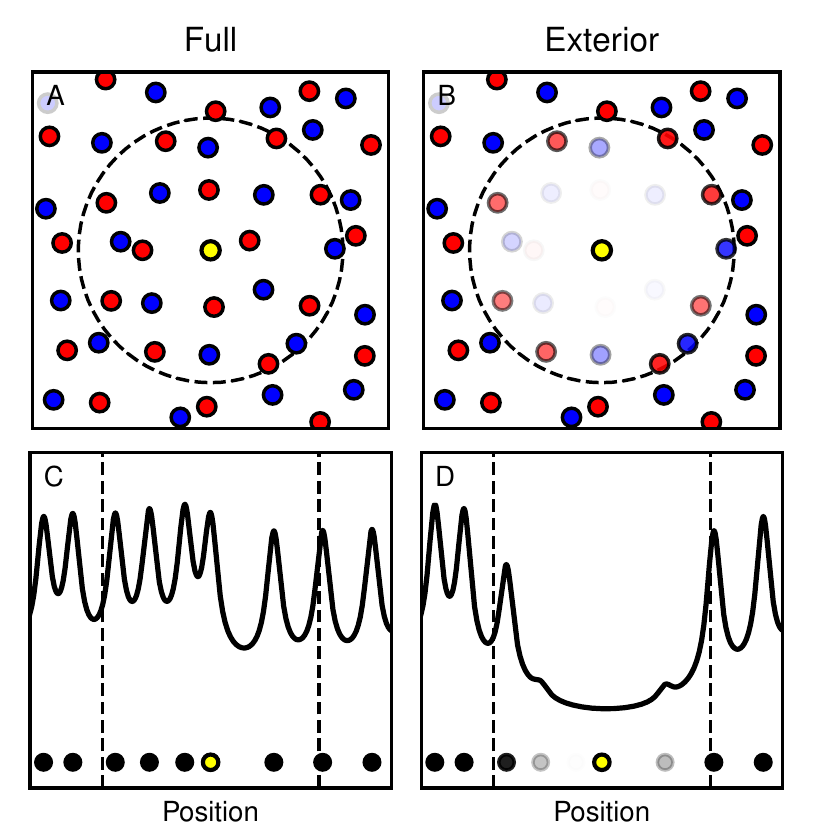}
  \caption{ Schematic representation of the atoms contributing to the full potential (A) and exterior potential features (B). Each circle represents an atom that contributes to the potential of the center atom shown in yellow, with the cutoff circle shown as a dotted line. In the latter case, the contribution of interior atoms is gradually turned on as their distance approaches the cutoff. Panels (C) and (D) show a one-dimensional example, where the vertical axis now shows the Coulomb potential (smeared out to remove the singularity) generated by the atoms. \rev{The large difference in the potential curve with and without} the interior contributions shows how the electrostatic
    potential around an atom is mostly dominated by its immediate neighbors, despite
    the long-range nature of the Coulomb potential.}
  \label{fig:intro}
\end{figure}

\subsection{Exterior potential features}\label{subsec_clean}

When computing physical interactions, the potential generated on a central atom $i$ by a set of 
$N$ atoms with weights $q_j$ at distances $r_{ij}$ can be expressed as
\begin{align}
  V_i^\mathrm{full} = \frac{1}{2}\sum_{j\neq i} q_j v(r_{ij}),
\end{align}
where $p > 0$ is the exponent associated with the interaction, e.g., \(p = 1\) for Coulomb and $p=6$ for dispersion forces.
The sum can be performed explicitly for a finite system, while a more general definition is necessary in the periodic case, as discussed in \autoref{sec:background}.
This atom-centered potential contains information on the position of atoms in the far-field. It is however not an ideal descriptor of long-range interactions: Despite being physically motivated, it also contains short-range effects from particles in the vicinity of the atom of interest.

In practice, short-ranged ML models are already excellent at extracting short-ranged
information from neighbor atoms up to a cutoff radius $\rcut>0$.
To realize ``range separated'', multi-scale models, it is therefore preferable to obtain  purely long-ranged features. 
\rev{We} introduce  ``exterior potential features'' (EPFs)
\begin{align}\label{potential_features_exterior}
  V_i^\mathrm{ext} & = \frac{1}{2}\sum_{r_{ij}} q_j f_\mathrm{trans}(r_{ij}) v(r_{ij}) \\
  & \approx \frac{1}{2}\sum_{r_{ij} >r_\mathrm{cut}} q_j v(r_{ij})
\end{align}
that eliminate the contributions from atoms within the cutoff, \rev{which} are already well-modeled by the short-range part of the model (see \autoref{fig:intro}A and B). The transition function $f_\mathrm{trans}$ essentially eliminates the contribution from interior atoms, but does so in a smooth way to ensure the differentiability of the features as atoms leave or enter the interior region. In the current implementation, it is explicitly given by
\begin{align}
    f_\mathrm{trans}(r) = \begin{cases}
        \left[ 1 - \cos\left(\frac{r^n}{\rcut^n}\right)\right] / 2 & r < \rcut \\
        0 & r \geq \rcut,
    \end{cases}
\end{align}
with some exponent $n$ that controls how quickly the contributions of the interior atoms vanishes as they approach the cutoff.

To illustrate the benefit of the exterior variant, consider the electrostatic (Coulomb)
potential generated by a one-dimensional chain of Gaussian charge distributions.
\autoref{fig:intro}C shows the construction leading to the full features, while \autoref{fig:intro}D corresponds to the exterior
variant. The stark difference in the two curves around the location of the center atom
$i$ shows that even for the very slowly decaying Coulomb potential, the potential around
an atom is most strongly influenced by the immediate neighbors. If the goal is to obtain
long-ranged features to combine them with a short-ranged model, we can see that the
exterior variant offers a clean signal, while the full potential features are
contaminated by the redundant  short-range contributions.
\torchpme provides an implementation of this idea, similar to a construction that has already been used in our previous
work~\cite{huguenin-dumittan_physics-inspired_2023}. This is a key difference with approaches that are based on the explicit modeling of physical interactions. 
In the following paragraphs, all examples use standard full potentials unless stated otherwise. Additionally, every time EPFs are mentioned, they are always used in the form that incorporates the smooth $f_\mathrm{trans}$ function.

\subsection{Implementing range separated ML architectures}

There are two main approaches one can take to incorporate long-range terms in an ML
architecture. One involves re-purposing traditional range separated calculators to
evaluate potential-like values at the atomic positions, and combine them with
short-range model architectures to compute short-range interactions, atomic
pseudo-charges, or arbitrary non-linear combinations of local descriptors and
atom-centered potentials. Given that the atom-centered potential values are (in the
limit of a converged mesh) translationally and rotationally invariant, this approach can
be easily applied to make invariant or equivariant predictions. In addition, a full
separation of a short and long-range model allows using multiple time stepping schemes to further reduce the overhead
stemming from the long-range evaluation~\cite{tuckerman_molecular_1991, kapil_accurate_2016}.

The second approach exploits the fact that mesh-based range separated calculators return
the potential values on a grid that covers the entire system. This makes it possible to
evaluate equivariant atom-centered features (along the lines of the LODE
framework~\cite{grisafi_incorporating_2019,huguenin-dumittan_physics-inspired_2023}) and
combine them with equivariant local descriptors within an equivariant architecture.

Even though the latter approach is more general (and can be used, for instance, to describe
many-body non-covalent interactions~\cite{huguenin-dumittan_physics-inspired_2023})
analytical considerations indicate that the most important information on long-range
pair interactions is contained in the invariant part of the descriptors (see also the SI
for a more thorough discussion). For this reason, the core components of the library we
developed focus on the evaluation of invariant features, which facilitate constructing
fast, flexible and streamlined range separated models. Nevertheless, the \torchpme API
does allow to evaluate fully equivariant LODE features quite easily, as we show in
\autoref{sub:lode}.

%% file: main/sections/application.tex
\section{Benchmarks and Model Architectures}\label{sec:results}

Having discussed some of the general principles, and the overall structure of our implementation, we now move on to discuss some practical examples and benchmarks.
If not
stated otherwise, all presented results are obtained using the \torchpme
implementation \rev{with single precision (32 bit) floating point arithmetics.} In \autoref{subsec_speed_and_accuracy} we begin by
presenting benchmarks on the speed and accuracy of the implementation. 
To further
validate the correct behavior of the ``full potential'' for more complex systems,
\autoref{subsec_md} presents MD trajectories in which our algorithms have been used to
compute the electrostatic forces via a LAMMPS interface, and compare the results against
LAMMPS's own P3M implementation. In order to showcase the flexibility of \torchpme, \autoref{subsec_learnability} illustrates how
our implementations easily allows learning different interaction types and charges.
Beyond these toy systems, \autoref{subsec_architectures} provides an example of a more
involved model architecture, and compares the results to previous work. Finally, \autoref{sub:lode} discusses how to use the building blocks of \torchpme to evaluate more complicated equivariant features.

\subsection{Benchmark: Speed and Accuracy}\label{subsec_speed_and_accuracy}
\begin{figure}
  \includegraphics{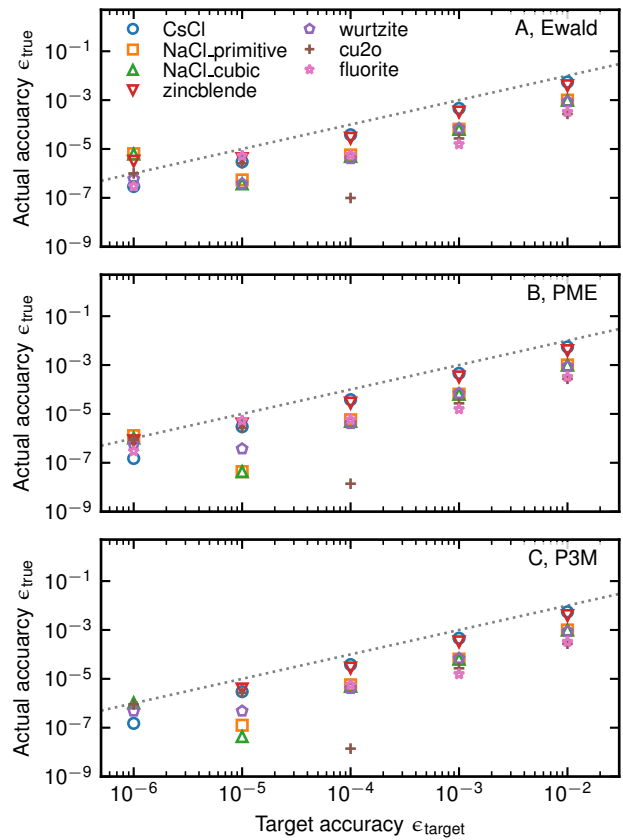}
  \caption{\label{fig:tuning}
    Empirical versus target accuracy after tuning calculator parameters for different
    replicated crystal structures and long-range methods using single precision floating
    point numbers. Panels show results for Ewald (A), PME (B) and P3M (C). Each crystal
    was replicated 32 times in each direction from its initial unit cell structure until. Different colored symbols represent different crystals. The true target
    accuracies are obtained from reference Madelung constants \cite{house_chapter_2020, glasser_solid-state_2012}.}
\end{figure}

\begin{figure*}
  \includegraphics[width=\textwidth]{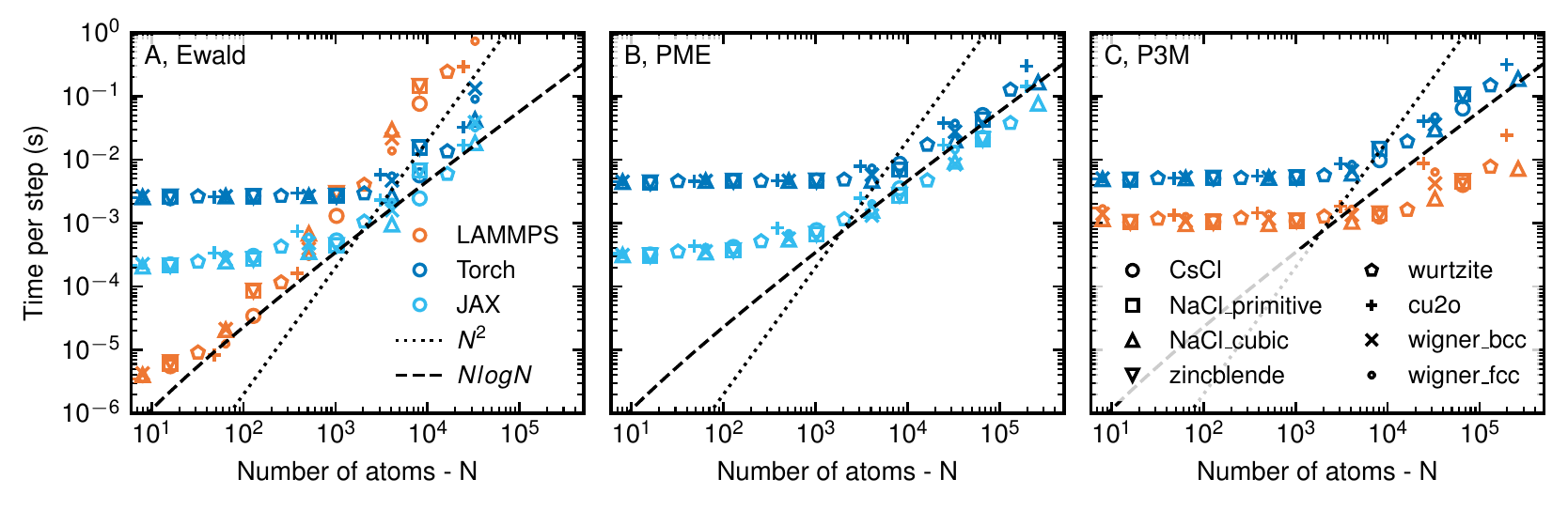}
  \caption{\label{fig:benchmark}
    Benchmark for computational cost of long-range calculators to run a single point evaluation of the
    energy, force and stress for different replicated crystal structures. Panels show results for Ewald (A)
    PME (B) and P3M (C). Different crystals are shown as different symbols. Each
    point is an average of 50 single point calculations of the same positions with a
  tuning accuracy of $1 \cdot10^{-4}$. Calculations for \torchpme and \jaxpme were performed with single precision (32-bit) arithmetics.}
\end{figure*}

We begin by discussing the results on the automatic hyperparameter tuning, as this also
determines the parameters used for the remaining tests on speed and accuracy. For many
simple crystal structures, lattice energies assuming purely electrostatic interactions
and ideal formal charges (commonly referred to as Madelung constants) are known and
tabulated with great accuracy. We define the actual (relative) accuracy as
\begin{equation}
    \epsilon_\mathrm{true} = \frac{|E_\mathrm{actual}-E_\mathrm{reference}|}{E_\mathrm{reference}} \, ,
\end{equation}
where
$E_\mathrm{actual}$ is the obtained energy and $E_\mathrm{reference}$ are the tabulated
reference energies. \autoref{fig:tuning} show the relative accuracy of the predicted
energies after tuning the parameters for the Ewald summation and the mesh calculators in
single precision (32-bit) floating point arithmetics for a given desired accuracy $\epsilon_\mathrm{desired} = \Delta F /
F_{qq}$, where $F_{qq}$ is the constant force created by two elementary charges
separated by $1$\AA{} to convert absolute errors into an estimated relative error as is also done in LAMMPS. We compute and show results for a few high-symmetry crystal structures with \rev{tabulated Madelung constants}\cite{house_chapter_2020, glasser_solid-state_2012}. The unit cell was
replicated 32 times in each direction, so that each structure contains at least 10,000
atoms. For each structure, the actual accuracy $\epsilon_\mathrm{true}$ is plotted
against the desired accuracy $\epsilon_\mathrm{target}$, meaning that ideally, all
points should lie below the black dotted line defined by $\epsilon_\mathrm{true} =
\epsilon_\mathrm{desired}$. Computation were performed on a single AMD EPYC 9334 CPU and one NVIDIA H100 SXM5 GPU showing. In Figure S6 in the supplementary material we show that a very similar behavior is observed for double-precision arithmetics.

For a diverse dataset containing structures with different numbers of atoms, we
recommend tuning the parameters on the system with the largest cell and largest number
of particles. A more detailed discussion of these considerations, as well as more detailed plots on the auto-tuning of the parameters and the convergence can be found in the supporting information.

Accuracy is only one aspect of a successful implementation. In
\autoref{fig:benchmark}, we demonstrate that \torchpme and \jaxpme are also competitive in
terms of computational cost by comparing them against LAMMPS's Ewald and P3M
implementations using the same replicated crystal structures shown in
\autoref{fig:tuning}. The LAMMPS benchmarks were performed using the version from August
2, 2023 (Update 3), in combination with the Kokkos 4.3.1 package. For our Ewald and P3M
benchmarks, we used the same parameters optimized by LAMMPS, \rev{to ensure a fair
comparison of the implementation. For a relative force accuracy of $10^{-4}$ the
LAMMPS parameters differ by less than 10\% when comparing to parameters obtained by our own
tuning code.} For PME we tune a separate set of parameters, aiming for an absolute force
accuracy of $10^{-4}$. In all cases, we employed a fixed cutoff of $4.4$\,Å, which
provided the best performance for our largest structure, containing
approximately $263'000$ atoms in a cubic cell with a side length of $65$\,Å (see
Figure S8 for a visualization of the grid search used to determine
this cutoff value). All reported timings include only the evaluation of the short- and
long-range components of the Coulomb potential, as well as its derivatives. Notably,
they do not account for the computation of the neighbor list. This exclusion is
justified because, in typical machine learning applications, the long-range part of the
architecture is often combined with a short-range one, whose neighbor list can
usually be reused. Additionally, in MD applications, the neighbor list can be reused for
multiple timesteps.

\autoref{fig:benchmark} shows that for small system sizes up to about $N=1000$ atoms,
LAMMPS Ewald (\autoref{fig:benchmark}A) is the fastest implementation, followed \rev{first by the Ewald method in \jaxpme and then} the PME method in
\jaxpme, and finally \torchpme. We can see that the cost for our implementation is more or
less constant across this range, meaning that this is not due to an inefficiency in the
actual algorithm, but rather a general overhead required during the initialization. Even
though these overheads are hard to quantify, they are probably related to better CUDA
kernel initialization and dispatch to the GPU by \jax. After a transition region between
$N=10^3$ and $N=10^4$, we observe that our Ewald implementations outperform the LAMMPS
one, while all codes show the expected scaling of
$\mathcal{O}(N^2)$~\cite{allentildesley}. \rev{In principle, one can tune the Ewald
method to $\mathcal{O}(N^{3/2})$ scaling by setting $\sigma \propto N^{1/6}$ and deducing the real and Fourier space cutoffs
accordingly. We show this scaling in the supplementary material in
Figure S7. However, this scaling cannot be achieved in a usual
machine learning workflow where the cutoff is fixed by the short-range model.} The
mesh-based algorithms are competitive with
the LAMMPS P3M implementation, while also showing the correct scaling of $\mathcal{O}(N\log
N)$~\cite{allentildesley}. For a system size of $N=10^4$, the mesh-based algorithms are
already faster than the Ewald ones by a factor of about five, with the gap widening
quickly due to the $\mathcal{O}(N^2)$ asymptotic scaling of the latter. From these
observations, we conclude that the implementations are fast enough for most practical
applications. \rev{We further recommend using the Ewald method during training, which 
usually involves small structures, where Ewald is faster than PME.}
After training, and when running production simulations, one should switch to the
PME or P3M calculators to achieve the best performance for larger systems. Note that
switching between the Ewald and mesh-based (PME and P3M) calculators is unproblematic as both can be tuned to achieve
high, and consistent, accuracy as demonstrated in \autoref{fig:tuning}, and \rev{share} fully compatible APIs.
\rev{As discussed further in the supplementary material, section S6, none of the benchmarked implementations are able to fully utilize available hardware on a modern GPU. We therefore expect further performance optimization to yield significant improvements.}

\subsection{Molecular Dynamics}\label{subsec_md}

\begin{figure}
  \includegraphics{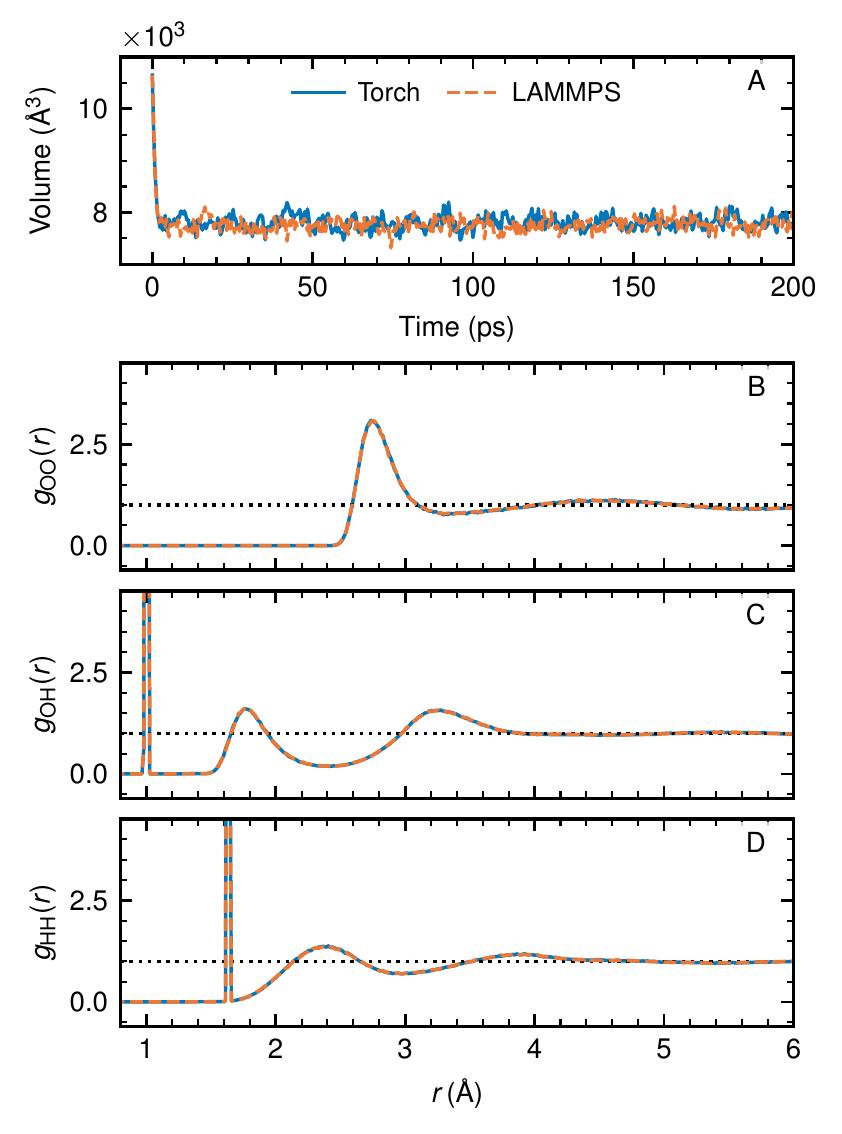}
  \caption{\label{fig:water}
    A:~Time series of the first $200\,\mathrm{ps}$ showing the fluctuating volume of an
    NpT simulation of rigid SPC/E water. The Coulombic part of the interactions are
    either handled by \rev{\torchpme's implementation of} PME (blue) or LAMMPS's P3M implementation (red). B -- D:
    Oxygen-oxygen $g_\mathrm{OO}$, oxygen-hydrogen $g_\mathrm{OH}$ and hydrogen-hydrogen
  $g_\mathrm{HH}$ radial distribution functions. \rev{The inset in B shows a snapshot of the initial configuration}}
\end{figure}

Even though \torchpme is designed with ML applications in mind, it is a good sanity
check to verify it can be used as an empirical force-field engine. To this end, we
construct a model that evaluates the electrostatic contributions for a rigid SPC/E water
model\cite{berendsen_missing_1987}, while the Lennard-Jones contributions are handled by
the simulation engine. For such a model, no training is performed since the charges for
the SPC/E model are fixed ($-0.8476\,\mathrm{e}$ for oxygen and $0.4238\,\mathrm{e}$ for
hydrogen). We use the Metatensor framework~\cite{metatensor} to export a
\emph{Torchscript} model that can be loaded and executed within the LAMMPS simulation
engine~\cite{thompson_lammps_2022}. We then perform a $2\,\mathrm{ns}$ NpT simulation of
$N=256$ rigid water molecules. The time step of the simulation is set to
$2\,\mathrm{fs}$ and the positions are recorded every $200$ steps ($0.2\,\mathrm{ps}$).
Temperature is controlled using the CSVR thermostat~\cite{bussi_canonical_2007} and the
pressure using a Nosé-Hoover barostat. During the simulation the Coulomb interaction
between the particles was either handled by LAMMPS or the \torchpme implementation, both
\rev{using hyperparameters whose (absolute) accuracy have been converged to $\epsilon =
2\cdot 10^{-5}\forceunits$.} Note that any of the grid-based implementations could also be used, but our
benchmarks show that, since the number of atoms is still small ($N<1000$), the Ewald
implementation is faster.
In \autoref{fig:water} (a) we show the time evolution of the system
volume, in which the cell equilibrates from the initial configuration (a cubic cell with
a side of 22~\AA), reaching the same equilibrium volume. From the volume fluctuations we
can compute the isothermal compressibility
\begin{align}
  \kappa_T = - \frac{1}{V} \left. \frac{\partial V}{\partial p} \right\vert_T.
\end{align}
\rev{The literature value for SPC/E water is $\kappa_\mathrm{T,ref} = (4.6 \pm
0.2)\cdot 10^{-10}\,\mathrm{m^2N^{-1}}$~\cite{vega_simulating_2011} which agrees well
with our result of $\kappa_\mathrm{T} = (4.8 \pm 0.3)
\cdot10^{-10}\,\mathrm{m^2N^{-1}}$ for \torchpme, and$\kappa_\mathrm{T,\mathrm{LAMMPS}} =
(4.7 \pm 0.1) \cdot10^{-10}\,\mathrm{m^2N^{-1}}$  for pure LAMMPS.} We obtain our values by leaving out
the first $10\,\mathrm{ps}$ of the simulation for equilibration and calculate the error
by a block average over 10 blocks. In \autoref{fig:water}B to D, we show the radial
distribution function between the two methods, which are essentially identical.

\begin{figure}[tbh]
\includegraphics{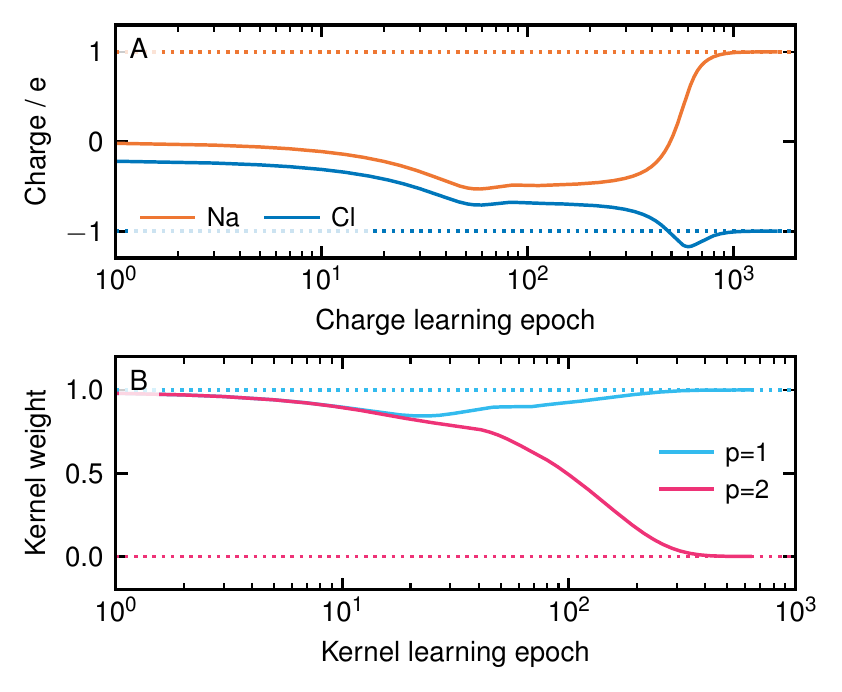}
\caption{\label{fig:learnability}
Training curves for the charge of the sodium and chloride atoms optimized on the energy of 10 random structures (A) and for optimizing the prefactor of two Fourier space kernels with different exponents $p$ (B). \rev{The inset in A shows one snapshot of 
the structures in the training set.}}
\end{figure}

\subsection{Learning charges and potentials}\label{subsec_learnability}

To illustrate how to use \torchpme as a flexible module for machine learning tasks, we design two toy examples that demonstrate the possibility of optimizing atomic pseudo-charges, and the functional form of the pair interactions. 
For both, we use a dataset of 1000 NaCl structures each consisting of
$N=64$ randomly placed atoms in cubic boxes of various side lengths that was already
used in previous work~\cite{grisafi_incorporating_2019,huguenin-dumittan_physics-inspired_2023}. Half of the atoms are Na atoms and
assigned a fixed charged of $q_\mathrm{Na}=+1$, while the other half are Cl atoms with a
charge of $q_\mathrm{Cl}=-1$. 
The total energy and forces of the systems are then
computed assuming a pure Coulomb interaction between the atoms with an external
reference calculator. 

In both of the models that we discuss, the optimization
was performed using the Adam optimizer with a maximum learning rate of $2\cdot10^{-1}$ as implemented in \torch.
In the first model, we make the atomic charges $q_\mathrm{Na}$ and $q_\mathrm{Cl}$ learnable
parameters, and initialize them to a small but non-zero value. Note that we do not explicitly
enforce charge neutrality, even though it could help for more complex systems. The two charges
are then optimized by fitting on the energy of the structures.
\autoref{fig:learnability} (a) shows the charges as a function of the number of epochs
in the training procedure. We can see that even though the convergence is not monotonic, the
charges reach the correct values after in fewer than $100$ epochs.

In the second model, we fix $q_\mathrm{Na}=+1$ and $q_\mathrm{Cl}=-1$ but this time
modify the interaction potential to be of the form
\begin{align}
  v(r|w_1,w_2) = \frac{w_1}{r} + \frac{w_2}{r^2}\,,
\end{align}
with learnable weights $w_1$ and $w_2$. \autoref{fig:learnability} (b) shows the
evolution of these weights, again as a function of the number of epochs. We can see that
after about $1000$ epochs, both weights converge to the expected, reference values of $w_1=1$ and
$w_2=0$.
Example files that were used to run these calculations are provided as examples in the
package repositories and documentation. 

\subsection{A model with learnable local charges}\label{subsec_architectures}

\begin{figure}
  \includegraphics{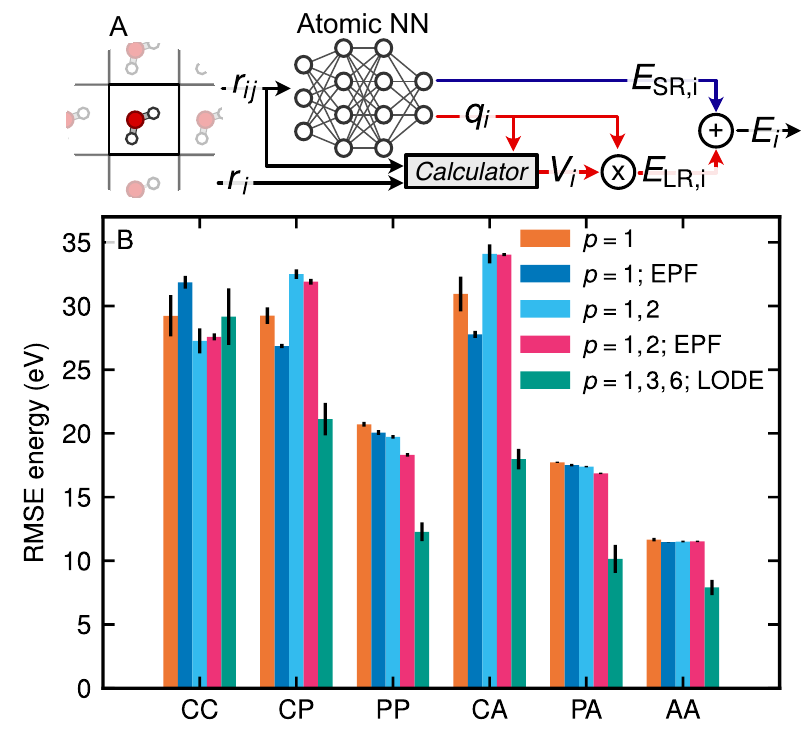}
  \caption{\label{fig:model_architecture}
    Schematic representation of an architecture that combines an atomic NN with a
    range separated calculator (A) and bar plots showing the comparison of RMSE on energies
    for different test subsets \rev{with error bars indicating the uncertainty estimated from 5 training runs} (B): charge-charge (CC), charge-polar (CP), polar-polar (PP),
    charge-apolar (CA), polar-apolar (PA), apolar-apolar (AA). Orange bars represent
    results from the architectures taken from
    Ref.~\citenum{huguenin-dumittan_physics-inspired_2023} while all other bars
    represent results from the architecture described in panel (A).}
\end{figure}

To demonstrate a more realistic application, we developed a complete end-to-end machine
learning model (\autoref{fig:model_architecture}A) and evaluated its performance on the organic molecules dataset used
in our previous work\cite{huguenin-dumittan_physics-inspired_2023} \rev{containing 2291 molecular pairs of 22 relaxed organic molecules in vacuum separated by various distances up to $15~\mathrm{\AA}$ in a periodic cell of $30~\mathrm{\AA}$}. \rev{This represents just one example among the many possible model architectures that can be constructed using \torchpme. This particular}  model has the overall structure of a ``third-generation'' neural network potential\cite{behl21cr}. 
It predicts the potential energy of a system,
which we decompose into atom-wise energy contributions: $E_\mathrm{system} =
\sum_{i=1}^{N} E_{i},$ where $E_\mathrm{system}$ is the total potential energy of the
system, $N$ is the number of atoms, and $E_{i}$ represents the energy contribution from
a single atom. We further decompose the atom-wise energy into short-range (SR) and
long-range (LR) components: $E_i = E_{\mathrm{SR},i} + E_{\mathrm{LR},i}$. The SR part is
computed using an atomic Neural Network (NN) which takes as input an invariant
descriptor representing the local atomic environment within a predefined cutoff radius
$r_\mathrm{cut}$: $E_\mathrm{SR, i} = f^\mathrm{SR}_{\theta^\mathrm{SR}}(\xi_i)$ where $\xi_i$ is a vector describing
the environment of atom $i$ and $\theta$ represents the learnable parameters of the function $f^\mathrm{SR}_\theta$. For our implementation, we used the Smooth Overlap of
Atomic Positions (SOAP) \cite{Csanyi2010} parameterization with the same parameters as
in Ref~\citenum{huguenin-dumittan_physics-inspired_2023}. The NN consists of three
layers, each with 256 neurons. Training was conducted for 200 epochs using the Adam
optimizer with the OneCycleLR policy\cite{smith2019super} and a maximum learning rate of
$5\cdot10^{-4}$. Besides predicting the $E_{\mathrm{SR},i}$ the same model predicts also
the atomic charges $q_i = f^\mathrm{charge}_{\theta^\mathrm{charge}}(\xi_i)$, which serve as inputs to a \torchpme
\emph{Calculator} to compute the long-range potential $V_i$ and energy $E_{\mathrm{LR},i}
= q_i \cdot V_i$. In \autoref{fig:model_architecture}B, we show a comparison of the
results obtained using our pipeline. Specifically, we consider four cases for $V_i$: (1)
using the full Coulomb potential ($p=1$), (2) using the Coulomb EPFs ($p=1$; EPF), (3)
using a linear combination of the Coulomb potential \rev{with a $1/r^3$ and $1/r^6$ term with learnable
weights ($p=1,3,6$)}, and (4) using \rev{the same linear combination as in (3) but with EPFs instead of the full potential features.} We compare our results to
those from Ref.~\citenum{huguenin-dumittan_physics-inspired_2023}, where LODE
descriptors were employed which includes on the order of $10^3$ features per atom,
whereas our current pipeline generates only a single descriptor. We find that we achieve
similar accuracy for the subset containing interactions between charged fragments, which is the dominating
contribution in the dataset. However, we tend to perform slightly worse on other
subsets, as these are no longer dominated by simple Coulomb-like interactions. This
discrepancy is unsurprising because the LODE descriptors use more features to effectively capture
other long-range interactions beside the Coulombic ones. Similar to what was
observed in Ref.~\citenum{huguenin-dumittan_physics-inspired_2023}, the inclusion of
additional exponents (e.g., \(p = 3\) and \(6\) in our case) does not yield better results.
Combined with learnable charges, a single descriptor restricted to long-range information is sufficient to capture the most prominent energy contributions, and that there isn't sufficient data to learn the weaker interactions with high accuracy. \rev{Note that the errors are largest for the CC pairs, since the Coulomb interaction is largest in absolute terms, consistently with our previous findings~\cite{huguenin-dumittan_physics-inspired_2023}. Further improvements would likely require more sophisticated architectures, which we are planning to explore in future work.} 
Finally, we observe that using EPFs provides only a marginal improvement in most cases (except for the charge-charge pair,
where a decrease in accuracy is expected due to the EPFs breaking strict physical
description). 
For this simple model and small dataset, \rev{whether or not} $E_\mathrm{LR}$ contains short-range contributions \rev{is not as relevant as these} can be compensated by the flexibility of $E_\mathrm{SR}$. 
A clear-cut range separation is still very useful -- e.g. to accelerate the evaluation of the model with multiple time stepping \cite{tuckerman_molecular_1991, kapil_accurate_2016} -- and it may prove beneficial in more complex models, or when used for long-range effects that are not dominated by electrostatics.

\begin{figure}
  \includegraphics[width=1.0\columnwidth]{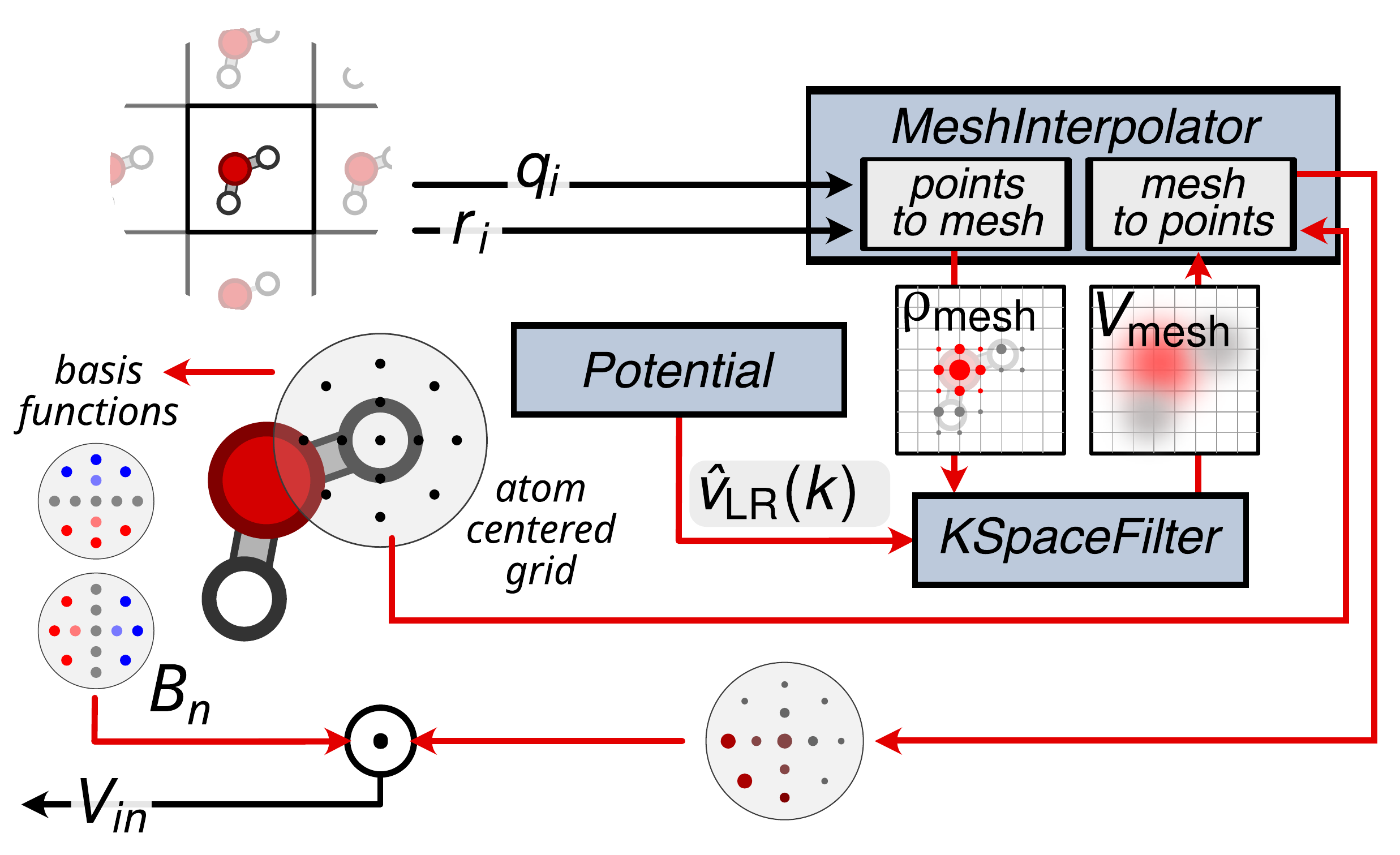}
  \caption{A schematic representation of the architecture of a model that computes LODE features, by evaluating the long-range part of the potential on a real-space grid, interpolating it on atom-centered grids, and combining it with a set of basis functions to evaluate atom-centered descriptors $V_{in}$, that are rotationally equivariant as long as the chosen basis functions are. 
  }
  \label{fig:lode-workflow}
\end{figure}

\subsection{Long distance equivariant features}
\label{sub:lode}

As an example of how modular design and automatic differentiation help adapt classical
algorithms for long-range interactions to modern machine learning tasks, we discuss
how to evaluate long-distance equivariant (LODE) features. \rev{A concise summary of the} basic idea behind
LODE\cite{grisafi_incorporating_2019}, and its generalization to $1/r^p$
potentials\cite{huguenin-dumittan_physics-inspired_2023}, \rev{is provided in Fig. 1 of Ref.~\citenum{huguenin-dumittan_physics-inspired_2023}. In essence, it consists of first computing} the potential  $V(\br) = \sum_i q_i
v(|\br-\mybold{r}_i|)$ generated by the atoms with weights $q_i$, and then to expand it in an atom-centered basis
\begin{equation}
  V_{in} = \int V(\br) B_n(\br-\mybold{r}_i)  \,\mathrm{d}\br\,,
  \label{eq:lode-integral}
\end{equation}
where $B_n$ are a set of basis functions. Usually, the basis functions are factorized in radial functions and spherical harmonics,
but this is inconsequential from the point of view of the implementation (see Fig.~\ref{fig:lode-workflow}).

The first step involves computing $V(\br)$ on a real-space grid. This can be
achieved with a Fourier convolution, exactly as in a grid-based long-range calculator. A
subtlety is that it is less convenient to include a short-range correction, because one
would have to evaluate it at all grid points, and not only at atomic positions,
requiring one to build an additional neighbor list and to compute a much larger number of
real-space pair interactions. Alternatively, and given that the objective is to build
long-range \emph{descriptors} rather than to compute a precise physical interaction, one
can exclusively use $v_\mathrm{LR}$ and compute only the reciprocal space component. If
one wants to do this while also subtracting the interior contribution, the cutoff
function has to be included in the computation of $\hat{v}_\mathrm{LR}$. Given that this
cannot be done analytically for a general, smooth cutoff function, we implement a
potential based on real-space cubic splines, which is numerically processed to generate
the Fourier kernel.

After having obtained $V(\br)$ on a regular mesh, we have to evaluate the
integral~\eqref{eq:lode-integral}. To do so, one can define an atom-centered mesh within
a sphere with the desired radius, and interpolate the potential at the grid points. Then,
the integral can be performed as a dot product with the basis function, pre-computed on
the same grid, and appropriate quadrature weights. Given that the
\emph{MeshInterpolator} class allows interpolating on arbitrary points, this
architecture can be realized very easily, and the resulting method is quite fast given
that the basis functions are pre-computed at initialization, competitive with the existing implementations of LODE~\cite{featomic} also for systems of moderate size.

%% file: main/sections/conclusion.tex
\section{Conclusions}

We have presented a highly flexible framework to perform
long-range ML efficiently, and provide two reference implementations in \torch and \jax.
Our method allows us to compute not only electrostatic and more general inverse
power-law potentials of the form $1/r^p$, but also to produce purified descriptors that
exclusively contain information about atoms outside of the cutoff radius.
This is useful when combining these features with short-range ML models, and can be used
to predict any invariant or equivariant target property. The features can be computed
using a variety of methods: Ewald summation is well-suited for
small to moderate system sizes with hundreds of atoms, while the particle-mesh algorithms such
as PME and P3M scale almost linearly with system size, $\mathcal{O}(N\log N)$, and are
hence well-suited for larger structures. \rev{All calculators fully support arbitrary unit cells, from cubic to triclinic ones. We provide a real space direct calculator for free boundary conditions systems and also plan extending our libraries using algorithms for 1D and 2D periodic systems \cite{neelov_particle-particle_2007, arnold_mmm1d_2005, arnold_mmm2d_2002, ballenegger_simulations_2009}.}

Our reference implementation fully exploit the flexibility and automatic differentiation capababilities of
the \torch and \jax libraries, allowing for smooth integration into any pre-existing ML
architecture. \rev{We also provide functions to automatically tune the numerical parameters of the methods,} further streamlining ease of use, with benchmarks confirming
the competitive speed and accuracy of the implementations. Besides the direct
application to improve existing short-range models, our implementations also allow an
easier inclusion of applied external potentials or fields into machine learning models~\cite{grisafi_symmetry-adapted_2018, ahrens-iwers_constant_2021}. The modular structure
also allows for various modifications for power-users who wish to fully exploit the
flexibility of the implementations. We hope that the availability of these libraries will encourage the development
of more standardized, efficient and scalable long-range ML models, and that their
modular design will allow to develop more flexible, general models than the simple
point-charge examples we discuss here.


%% file: main/sections/supplementary.tex
\section{Supplementary Material}
The electronic supporting material for this publication include further details on the
implementation, different models and datasets, and additional results on benchmarks. The
source code \torchpme and \jaxpme packages are publicly available at
\url{https://github.com/lab-cosmo/torch-pme} and
\url{https://github.com/lab-cosmo/jax-pme}. Some results are compared with the
generalized LODE descriptors introduced in previous work by the authors, which have been
computed using the \emph{featomic} (formerly \emph{rascaline}) package \cite{featomic}, available at
\url{https://github.com/metatensor/featomic}.

%% file: main/sections/acknowledgments.tex
\section*{Acknowledgments}

MC, KKHD, MFL, PL and WBH acknowledge funding from the European Research Council (ERC) under the research and innovation programme (Grant Agreement No. 101001890-FIAMMA). MC, PL, ER and MH acknowledge funding from the NCCR
MARVEL, funded by the Swiss National Science Foundation (SNSF, grant 
number 182892), which also supported a research internship for MH through an Inspire Potentials Fellowship.
MFL acknowledges funding from the German Research Foundation (DFG) under project number 544947822.
We would like to thank the members of the Laboratory of Computational Science and Modeling, and especially Guillaume Fraux, Filippo Bigi and Arslan Mazitov, for their contributions to the software infrastructure that enabled this study.
We further thank Philipp Stärk for useful discussions.

%% file: main/sections/data_availability.tex
\section*{Data Availability Statement}
The source code to our \torch and \jax libraries are available on Github at
\url{https://github.com/lab-cosmo/torch-pme} and
\url{https://github.com/lab-cosmo/jax-pme}. All presented results have been computed using version 0.2 of \torchpme and commit db1e662bdd8e8a68f675b05a94693e39dafc0fdd of \jaxpme. Code to reproduce \autoref{fig:learnability}
as well as how to construct LODE densities are available as tutorials in the \torchpme
documentation. The datasets used in this work are taken from our previous work~\cite{huguenin-dumittan_physics-inspired_2023} and can either be found in its supporting information, or more directly at the \emph{materialscloud} database entry \url{https://archive.materialscloud.org/record/2023.151}. All other codes and input files for computing the accuracies, running
benchmarks, MD simulations, and the model with learnable local charges are as well
available on Zenodo at \url{https://zenodo.org/records/14792591}.

%% file: si/sections/nomenclature.tex
\nomenclature[N]{\(\mathbb{N}\)}{Natural Numbers $1,2,\dots$}
\nomenclature[N]{\(\mathbb{Z}\)}{Integers} \nomenclature[N]{\(\mathbb{R}\)}{Real
numbers} \nomenclature[N]{\(\mathbb{R}^3\)}{3-dimensional Euclidean space}
\nomenclature[N]{\(\mathbb{C}\)}{Complex numbers}

\nomenclature[P]{\(\Gamma\)}{Lattice}
\nomenclature[P]{\(\Gamma^*\)}{Reciprocal Lattice}
\nomenclature[P]{\(\ba_j\)}{Basis vectors $\ba_1,\ba_2,\ba_3$ of lattice}
\nomenclature[P]{\(\bb_j\)}{Basis vectors $\bb_1,\bb_2,\bb_3$ of reciprocal lattice}
\nomenclature[P]{\(\br\)}{General position vector (element of $\br\in\real^3$)}
\nomenclature[P]{\(\br_i\)}{Position vector of atom $i$}
\nomenclature[P]{\(r\)}{Standard norm (2-norm) or magnitude of vector $\br$}
\nomenclature[P]{\(r_{ij}\)}{Distance between atoms $i$ and $j$}
\nomenclature[P]{\(\boldsymbol{\Omega}\)}{Unit cell}
\nomenclature[P]{\(\Omega\)}{Volume of Unit cell}
\nomenclature[P]{\(\bk\)}{Reciprocal space vector either in $\bk\in\real^3$ or a
reciprocal lattice $\bk\in\Gamma^*$}
\nomenclature[P]{\(k\)}{Standard norm (2-norm) or magnitude of vector $\bk$}
\nomenclature[P]{\(\Gamma^*\)}{Reciprocal Lattice}
\nomenclature[P]{\(\bl\)}{Lattice vector $\bl\in\Gamma$}
\nomenclature[P]{\(f\)}{Generic function (potentially in a suitable function space)}
\nomenclature[P]{\(\hat{f}\)}{Fourier transform of function $f$}

\nomenclature[I]{\(r_\mathrm{cut}\)}{Cutoff radius in real space}
\nomenclature[I]{\(k_\mathrm{cut}\)}{Cutoff radius in reciprocal space, related to
long-range resolution via $k_\mathrm{cut} = \frac{2\pi}{h}$}
\nomenclature[I]{\(h\)}{Long-Range Resolution, related to reciprocal space cutoff via
$k_\mathrm{cut} = \frac{2\pi}{h}$} \nomenclature[I]{\(\sigma\)}{Smearing parameter
determining the transition between short-range and long-range functions}
\nomenclature[I]{\(V_\mathrm{bare}\)}{Bare potential, usually taken to be of the inverse
  power-law form $V_\mathrm{bare}(r) = \frac{1}{r^p}$ with the Coulomb potential as the
special case for $p=1$} \nomenclature[I]{\(p\)}{Exponent in the bare potential
$V_\mathrm{bare}$} \nomenclature[I]{\(V_\mathrm{SR}\)}{Short-range part of the bare
potential} \nomenclature[I]{\(V_\mathrm{LR}\)}{Long-range part of the bare potential}
\nomenclature[I]{\(\hat{V}_\mathrm{LR}\)}{Fourier transform of the long-range part of
the bare potential} \nomenclature[I]{\(V_i\)}{Potential at the location of atom $i$
(typically excluding the contributions by atom $i$ itself)}
\nomenclature[I]{\(V_\mathrm{charge}\)}{A position-independent term required to
compensate in case the unit cell has a net charge} \nomenclature[I]{\(q_i\)}{Charge of
atom $i$, or more generally for non-Coulombic potentials, the weight of atom $i$}

\printnomenclature{}

%% file: si/sections/conventions.tex
\section{Mathematical Conventions, Notation and Prerequisites}\label{sec_conventions}
Many different conventions and notations can be found in the literature for mathematical
expressions and concepts. Most relevant to this work are the ones relating to periodic lattices and Fourier transforms. To keep a consistent notation and
convention throughout this document, we summarize all definitions in this section. We also briefly summarize some well-known mathematical results about these that are used throughout this document and the implementation.

\subsection{Vectors, Norms and Inner Products}\label{subsec_vectors} Throughout this
document, we shall use the bold-font $\br\in\real^3$ for vectors in Euclidean space,
while $r=\|\br\|$ will denote its magnitude computed using the usual 2-norm. Similarly,
we will use $\bk\in\real^3$ and $k=\|\bk\|$ when discussing Fourier transforms, and
write either $\bk\cdot \br$ or just $\bk\br$ for the inner product between two vectors.

For an atomic system consisting of $N\in\mathbb{N}$ atoms, we index the atoms by
$i,j\in\lbrace 1,2,\dots,N\rbrace$. This is then used to write $\br_i$ for the position
vector of atom $i$,  and $r_{ij} = \|\br_j - \br_i\|$ for the distance between atoms $i$
and $j$.

For the more mathematically inclined readers: $\br$ and $\bk$ technically are elements
of different vector spaces, both isomorphic to $\real^3$, which are duals of each other.
In practice, the choice of symbol will unambiguously determine which space is meant.

\subsection{Lattice and Reciprocal Lattice}\label{subsec_lattice} The lattice $\Gamma$
spanned by the three linearly independent vectors $\ba_1,\ba_2,\ba_3\in\real^3$ is
defined as
\begin{align}
  \Gamma & = \mathrm{span}_\mathbb{Z}\lbrace \ba_1,\ba_2,\ba_3 \rbrace \\
  & = \lbrace \boldsymbol{l} =
  n_1\ba_1 + n_2\ba_2 + n_3\ba_3 | n_1,n_2,n_3\in\mathbb{Z}\rbrace.
\end{align}
In practice, these basis vectors will be chosen to specify the periodicity within the
cell. We can then define the reciprocal lattice $\Gamma^*$ by the requirement
\begin{align}
  \Gamma^* = \lbrace \bk \in \real^3| \forall \br \in \Gamma: \bk\br \in 2\pi\mathbb{Z} \rbrace
\end{align}
which can be shown to be the lattice spanned by the three basis vectors
\begin{align}
  \bb_i = 2\pi \frac{\ba_j \times \ba_k}{\ba_i \cdot (\ba_j \times \ba_k)},
\end{align}
where $(i,j,k)$ is an even permutation of $(1,2,3)$ and $\times$ is the cross-product.
We will generally use the symbol $\boldsymbol{l}\in\Gamma$ to denote an element of the
(real space) lattice, i.e. a different symbol from $\br$, while we use the same symbol
$\bk\in\Gamma^*$. It should be clear from context which one is meant. In particular, we
will write
\begin{align}
  \sum_{\bk}
\end{align}
as a shorthand for
\begin{align}
  \sum_{\bk\in\Gamma^*}
\end{align}
as the former is likely to be more familiar to be many readers. The summation symbol is
emphasizing that $\bk$ only runs over the discrete set $\Gamma^*$ rather than $\real^3$.

Note that we do include the factor of $2\pi$ directly in the definition of the
reciprocal space vectors, since this leads to slightly more compact formulae and also is
slightly more convenient for implementations.

\subsection{Unit cells and volumes}\label{subsec_unitcell} In practice, the basis
vectors $\ba_1,\ba_2,\ba_3$ introduced in the previous subsection will be used to
specify a periodic repetition of atoms. In this case, the volume of the unit cell
spanned by the three vectors can be obtained by computing their triple product, and will
be denoted by
\begin{align}
  \Omega = \ba_1 \cdot (\ba_2 \times \ba_3).
\end{align}

In order to compute Fourier transforms, it will also be necessary to integrate over such
a unit cell. We can thus define the closed set
\begin{align}
  \boldsymbol{\Omega} = \lbrace w_1\ba_1 + w_2\ba_2 + w_3\ba_3 | w_1,w_2,w_3\in[0,1] \rbrace
\end{align}
for the unit cell, such that its volume $\Omega = \mathrm{Vol}(\boldsymbol{\Omega})$.

For the mathematically inclined readers: strictly speaking, it would be more correct to
define $\boldsymbol{\Omega} = \real^3 / \Gamma$ using a quotient, where $\Gamma$ is the
lattice defined by the basis vectors. With this definition, the periodic structure of
the underlying domain becomes more explicit, and it is easy to see that
$\boldsymbol{\Omega}$ is diffeomorphic to the 3-torus $\mathbb{T}^3 = S^1 \times S^1
\times S^1$ as a smooth manifold.

\subsection{Fourier Series and Fourier Transforms}\label{subsec_fourier} In this
chapter, we shall make extensive use of the Fourier transform in its various forms.

Firstly, consider functions $f:\real^3\rightarrow\mathbb{C}$ (or more precisely,
absolutely integrable functions $f\in L^1(\real^3,\mathbb{C})$). Then, for all
$\bk\in\real^3$, we define the Fourier transform of $f$ as
\begin{align}\label{fourier_tf_l1}
  \hat{f}(\bk) = \int_{\real^3}\mathrm{d}^3r\, e^{-i\bk\br}f(\br).
\end{align}
The inverse transformation then is given by
\begin{align}\label{inverse_tf_l1}
  f(\br) = \frac{1}{(2\pi)^3}\int_{\real^3}\mathrm{d}^3k\, e^{i\bk \br}\hat{f}(\bk).
\end{align}
The entire factor of $\frac{1}{(2\pi)^3}$ is thus contained in the inverse transform.
While this is arguably less elegant from the point of view of symmetry, it will lead to
slightly simpler expressions when comparing this to the Fourier series for periodic
functions, which will be helpful when  discussing the Poisson summation formula.

If the original function $f$ is rotationally symmetric, often written as $f(\br) =
f(r)$, then so is its Fourier transform $\hat{f}(\bk) = \hat{f}(k)$. By doing the
integral in spherical coordinates and integrating with respect to the angular variables
$(\phi,\theta)$, we can obtain the more direct formula relating $f$ to $\hat{f}$ as the
one-dimensional integral
\begin{align}\label{fourier_tf_radial}
  \hat{f}(k) & = 4\pi \int_0^\infty \mathrm{d}r\, r^2 \frac{\sin(kr)}{kr}f(r) \\
  & = \frac{4\pi}{k} \int_0^\infty \mathrm{d}r\, r\sin(kr)f(r)
\end{align}
and its inverse
\begin{align}\label{inverse_tf_radial}
  f(r) & = \frac{1}{(2\pi)^3} \frac{4\pi}{r} \int_0^\infty \mathrm{d}k\, k\sin(kr)\hat{f}(k).
\end{align}
In particular, this means that apart from a factor of $\frac{1}{(2\pi)^3}$, the radially
symmetric transform is its own inverse. This version of the transform is closely related
to the Hankel transform.

So far, we have discussed the Fourier transform of a generic (absolutely integrable)
function on $\real^3$. We now move on to the Fourier series representation of a periodic
function.

Let $f:\real^3\rightarrow\mathbb{C}$ be a function having the periodicity of a unit cell
$\boldsymbol{\Omega}$ with its corresponding lattice $\Gamma$. More mathematically, this
means that for all $\br\in\real^3$ and $\bl\in\Gamma$, $f(\br+\bl) = f(\br)$. The
function is clearly specified by just its values in the unit cell $\boldsymbol{\Omega}$.

For any reciprocal space vector $\bk\in\Gamma^*$, we then define the Fourier transform
(for the periodic case also called Fourier series coefficient) of $f$ as
\begin{align}\label{fourier_tf_periodic}
  \hat{f}(\bk) = \int_{\boldsymbol{\Omega}}\mathrm{d}^3r\,e^{-i\bk\br}f(\br).
\end{align}
We can then recover $f$ using the inverse transform
\begin{align}\label{inverse_tf_periodic}
  f(\br) = \frac{1}{\Omega}\sum_{\bk \in \Gamma^*}\hat{f}(\bk)e^{i\bk\br}.
\end{align}

Note that a common convention is to put the factor of $1/\Omega$ in the (forwards)
Fourier transform rather than the inverse. Our convention makes the expressions for the
forward transforms for the aperiodic case in \autoref{fourier_tf_l1} and periodic one in
\autoref{fourier_tf_periodic} more symmetric.

\subsection{The Poisson Summation Formula}\label{subsec_poisson} The following theorem
provides us with an explicit connection between summation in real space and reciprocal
space.

\begin{theorem}[Poisson Summation Formula]
  Let $f\in L^1(\real^3,\mathbb{C})$ be an absolutely integrable function, $\Gamma$ a
  lattice whose unit cell $\boldsymbol{\Omega}$ has a volume $\Omega$ and $\Gamma^*$ its
  reciprocal lattice. Then, for all $\br \in \real^3$, we get
  \begin{align}
    \sum_{\bl \in \Gamma} f(\br + \bl) = \frac{1}{\Omega}\sum_{\bk \in \Gamma^*} \hat{f}(\bk)e^{i\bk\br}.
  \end{align}
  In other words, a sum of the values of $f$ over a lattice $\Gamma$ can be expressed as
  a sum over reciprocal space vectors.
\end{theorem}
While this is a standard result in mathematical methods, it is so important to our work
that we provide a short proof for the convenience of the reader. There is no originality
in the proof, and this is merely to make this document more self-contained.
\begin{proof}
  Let $F(\br) = \sum_{\bl \in \Gamma} f(\br + \bl)$ be the left-hand side. We will first
  show that $F$ is a periodic function with respect to the lattice $\Gamma$. From the
  inversion of Fourier transforms for periodic functions, this then means that we can
  write $F$ using its Fourier series coefficients $\hat{F}(\bk)$ as
  \begin{align}
    F(\br) = \frac{1}{\Omega} \sum_{\bk\in\Gamma^*}\hat{F}(\bk)e^{i\bk\br}.
  \end{align}

  Comparing this with the theorem, we can see that all we need to do is to show that
  $\hat{F}(\bk) = \hat{f}(\bk)$ for all $\bk\in\Gamma^*$.

  Starting with the periodicity of $F$, this comes from the fact that for two lattice
  vectors $\bl,\bl'\in\Gamma$, their sum is still just a lattice vector $\bl'' =
  \bl+\bl' \in\Gamma$. A sum over all $\bl'\in\Gamma$ is equivalent to a sum over all
  $\bl''=\bl+\bl'$, since we are effectively just shifting the origin of the lattice by
  $\bl$. Thus,
  \begin{align}
    F(\br + \bl) & = \sum_{\bl'\in\Gamma} f(\br + \bl + \bl') \\
    & = \sum_{\bl''\in\Gamma} f(\br + \bl'') \\
    & = F(\br).
  \end{align}

  We can thus compute the Fourier transform of $F$ (the variant for periodic functions,
  also called Fourier series) directly from the definition in
  \autoref{fourier_tf_periodic}. For $\bk\in\Gamma^*$, we get
  \begin{align}
    \hat{F}(\bk) & = \int_{\boldsymbol{\Omega}}\mathrm{d}^3r\, e^{-i\bk\br} F(\br) \\
    & = \int_{\boldsymbol{\Omega}}\mathrm{d}^3r\,e^{-i\bk\br}  \sum_{\bl \in \Gamma} f(\br + \bl) \\
    & \overset{Fubini}{=} \sum_{\bl \in \Gamma} \int_{\boldsymbol{\Omega}}\mathrm{d}^3r\, e^{-i\bk\br} f(\br + \bl) \\
    & \overset{\bk\bl=2\pi n}{=} \sum_{\bl \in \Gamma} \int_{\boldsymbol{\Omega}}\mathrm{d}^3r\,e^{-i\bk(\br+\bl)} f(\br + \bl) \\
    & \overset{\br'=\br+\bl}{=} \int_{\real^3}\mathrm{d}^3r'\, e^{-i\bk \br'}f(\br') \\
    & = \hat{f}(\bk).
  \end{align}
  This shows that the Fourier series coefficients $\hat{F}(\bk)$ of the periodic
  function $F$ are actually identical to the Fourier transform of the (in general,
  aperiodic) function $f$. The only key difference is that $\hat{f}(\bk)$ is defined for
  all $\bk\in\real^3$, while $\hat{F}(\bk)$ only is defined for $\bk\in\Gamma^*$. Thus,
  $\hat{F}$ is merely the restriction of $\hat{f}$ onto the reciprocal lattice
  $\Gamma^*$.

  We can now use the inverse transform for $F$, which allows us to write
  \begin{align}
    F(\br) = \frac{1}{\Omega} \sum_{\bk\in\Gamma^*}\hat{F}(\bk)e^{i\bk\br}.
  \end{align}
  But since $\hat{F}(\bk) = \hat{f}(\bk)$, this leads to
  \begin{align}
    F(\br) = \frac{1}{\Omega} \sum_{\bk\in\Gamma^*}\hat{f}(\bk)e^{i\bk\br}.
  \end{align}
  Remembering that we had defined $F(\br) = \sum_{\bl}f(\br+\bl)$, this concludes the
  theorem.
\end{proof}

%% file: si/sections/ewald_derivation.tex
\section{Derivation of Ewald Summation}\label{sec_ewald_derivation}
\subsection{General Expression using Poisson Summation}
In this section, we derive the general decomposition of the potential into its four terms in the Ewald summation. For an explanation of the notation and terminology related to unit cells, periodic lattices and Fourier transforms, please refer to the previous section.

The results in this section are already known in the literature. The purpose of this section is to make this paper self-contained, and to include derivations of all results using a consistent notation that we also use in the main text.

\subsection{Self-Term in Ewald Summation}
The self-term in Ewald summation can be obtained by computing the limit
\begin{align}
    v_\mathrm{self} & = \lim_{r\rightarrow 0} v_\mathrm{LR}(r).
\end{align}
To evaluate this, we can use the Taylor series expansion of the lower incomplete Gamma function, which can be derived directly from its definition and using Fubini's theorem to interchange the order of summation and integration for $a>0$:
\begin{align}
    \gamma(a,x) & = \int_0^x \mathrm{d}t\, t^{a-1}e^{-t} \\
    & = \int_0^x\mathrm{d}t\, t^{a-1}\sum_{n=0}^\infty \frac{(-1)^n}{n!} t^n \\
    & = \sum_{n=0}^\infty \frac{(-1)^n}{n!} \int_0^x \mathrm{d}t\, t^{n + a-1} \\
    & = \sum_{n=0}^\infty \frac{(-1)^n}{n!} \frac{1}{a+n} x^{a+n} \\
    & = \frac{1}{a}x^a + \mathcal{O}\left(x^{a+1}\right)\,.
\end{align}
Using this, we obtain
\begin{align}
    v_\mathrm{self} & = \lim_{r\rightarrow 0} \frac{1}{\Gamma\left(\frac{p}{2}\right)}\frac{\gamma\left(\frac{p}{2},\frac{r^2}{2\sigma^2}\right)}{r^p} \\
    & = \frac{1}{\Gamma\left(\frac{p}{2}\right)} \frac{2}{p} (2\sigma^2)^{-\frac{p}{2}}\\
    & = \frac{1}{\Gamma\left(\frac{p}{2}+1\right) (2\sigma^2)^{\frac{p}{2}}}
\end{align}

\subsection{Charge Neutralization term in Ewald Sum}
The correction term to enforce charge neutrality arises from the interaction of the particles with a homogeneous background charge density that is used to neutralize the system.

For this, consider the Green's function $G(\br,\br')$ describing the potential at position $\br$, assuming the presence of a particle at $\br'$ with weight (or charge) $q=1$, including its periodic images and a homogeneous background with total charge $-q$ per unit cell to fulfill charge neutrality. Then,
\begin{align}
    G(\br,\br') & = \sum_{\bl\in\Gamma} v(\br-\br'+\bl) - \frac{1}{\Omega} \int_{\real^3}\mathrm{d}\br'\, v(\br-\br') \\
    & = \sum_{\bl\in\Gamma} v_\mathrm{SR}(\br-\br'+\bl) + \sum_{\bl\in\Gamma} v_\mathrm{LR}(\br-\br'+\bl) - \frac{1}{\Omega} \int_{\real^3}\mathrm{d}\br'\, v(\br-\br') \\
    & = \sum_{\bl\in\Gamma} v_\mathrm{SR}(\br-\br'+\bl) + \frac{1}{\Omega} \sum_{\bk\neq 0} \hat{v}_\mathrm{LR}(\bk) e^{i\bk(\br-\br')} - \frac{1}{\Omega} \int_{\real^3}\mathrm{d}\br'\, v_\mathrm{SR}(\br-\br'),
\end{align}

where the Poisson summation formula was used in the last step. The first two terms simply correspond (once summing over neighbor atoms $j$) to the real and reciprocal space sums in Ewald summation. The last term is the desired correction term, which represents an additional term in the potential due to the homogeneous background charge.

To evaluate this last term, we define the convenient shorthand
\begin{align}
    v_\mathrm{charge} & = \int_{\real^3}\mathrm{d}\br\, v_\mathrm{SR}(\br).
\end{align}
The integrand is just written as $v_\mathrm{SR}(\br)$ instead of $v_\mathrm{SR}(\br-\br')$, as the integral runs over the entire space anyway. The actual charge correction term due to an atom with charge $q$, with its compensating background charge being spread over volume $\Omega$ and taking into account an additional factor of $\sfrac{1}{2}$ from double counting thus gives
\begin{align}
    V_{i,\mathrm{charge}} = -\frac{1}{2}\frac{q}{\Omega}.
\end{align}

The calculation of the charge correction term has now been reduced to the computation of
\begin{align}
    v_\mathrm{charge} & = \int_{\real^3}\mathrm{d}\br\, v_\mathrm{SR}(r) \\
    & = 4\pi \int_0^\infty \mathrm{d}r\,r^2 v_\mathrm{SR}(r),
\end{align}
where we used spherical symmetry in the last step.

While we could directly start computing this for the general case, it is instructive to first study the behavior at the two bounds of the integral, i.e. what happens in the limits as $r\rightarrow 0$ and $r\rightarrow\infty$. Firstly, since $v_\mathrm{SR}(r)$ was explicitly constructed to decay quickly as $r\rightarrow \infty$, the upper bound of integration does not cause any problems. On the other hand, we know that $v_\mathrm{SR}$ contains the singularity of the bare $1/r^p$ potential as $r\rightarrow 0$. To estimate the lower bound of integration, we can thus study the simplified integral
\begin{align}
    \int_0^R \mathrm{d}r\,r^2 \frac{1}{r^p} & = \int_0^R \mathrm{d}r\, r^{2-p}, \\
    & = \begin{cases}
        \frac{1}{3-p}r^{3-p} \Big |_0^R & p \neq 3 \\
        \log p \Big |_0^R & p = 3 
    \end{cases} \\
    & = \lim_{\epsilon\rightarrow 0}\begin{cases}
        \frac{1}{3-p}R^{3-p} - \epsilon^{3-p} & p < 3 \\
        \log R - \log \epsilon & p = 3 \\
        -\frac{1}{p-3}\frac{1}{R^{p-3}} + \frac{1}{p-3} \frac{1}{\epsilon^{p-3}}& p > 3 \\
    \end{cases}
    \\
    & = \begin{cases}
        \frac{1}{3-p}R^{3-p} & p < 3 \\
        +\infty & p = 3 \\
        +\infty & p > 3
    \end{cases}
\end{align}
Note that for the third case where $p>3$, we wrote all results in terms of $p-3 > 0$ instead of $3-p < 0$ to make all prefactors and exponents positive. These results clearly show that due to the strong singularity at zero distance, the charge correction term due to a homogeneous background can only be evaluated for $p<3$.

Now focusing only on $p<3$ where the full integral for $v_\mathrm{charge}$ is well-defined, we obtain
\begin{align}
    v_\mathrm{charge} & = 4\pi \int_0^\infty \mathrm{d}r\,r^2 v_\mathrm{SR}(r) \\
    & = \frac{4\pi }{\Gamma\left(\frac{p}{2}\right)} \int_0^\infty \mathrm{d}r\,r^{2-p} \Gamma\left(\frac{p}{2},\frac{r^2}{2\sigma^2}\right) \\
    & = \frac{4\pi }{\Gamma\left(\frac{p}{2}\right)} \int_0^\infty \mathrm{d}r\,r^{2-p} \int_{\frac{r^2}{2\sigma^2}}^\infty \mathrm{d}t\, t^{\frac{p}{2}-1}e^{-t}
\end{align}
Using Fubini's theorem to swap the order of the two integrals and appropriately adjusting the integral boundaries (which is allowed for $p<3$ since the integrand can be estimated by a power law), we obtain
\begin{align}
    v_\mathrm{charge} & = \frac{4\pi}{\Gamma\left(\frac{p}{2}\right)} \int_{0}^\infty \mathrm{d}t\, t^{\frac{p}{2}-1}e^{-t} \int_0^{\sqrt{2\sigma^2 t}}\mathrm{d}r\,r^{2-p} \\
    & = \frac{4\pi}{\Gamma\left(\frac{p}{2}\right)} \int_{0}^\infty \mathrm{d}t\, t^{\frac{p}{2}-1}e^{-t} \frac{1}{3-p}\left(2\sigma^2 t\right)^{\frac{3-p}{2}} \\
    & = \frac{4\pi \left(2\sigma^2 \right)^{\frac{3-p}{2}} }{(3-p)\Gamma\left(\frac{p}{2}\right)} \int_{0}^\infty \mathrm{d}t\, t^{\frac{1}{2}}e^{-t} \\
    & = \frac{2\pi^\frac{3}{2} \left(2\sigma^2 \right)^{\frac{3-p}{2}} }{(3-p)\Gamma\left(\frac{p}{2}\right)}.
\end{align}

For the two special cases of $p=1$, this reduces to
\begin{align}
    v_\mathrm{charge}^{p=1} = 2\pi \sigma^2
\end{align}
and for $p=2$ to
\begin{align}
    v_\mathrm{charge}^{p=2} =  (\sqrt{2}\pi)^3\sigma = 2^{\frac{3}{2}}\pi^3\sigma
\end{align}

\subsection{Treatment of Charge Compensation Term}
Summarizing the results on the compensation term and the subtlety regarding the critical exponent of $p=3$:
\begin{enumerate}
    \item For $p\leq 3$, the sums used to compute the potential are divergent if the cell is not charge neutral, thus requiring the addition of a homogeneous charge density of opposite total charge to ensure a neutral cell. The potential of some particle $i$ then consists of contributions from the remaining particles (including periodic images), and a second term $V_{i,\mathrm{charge}}$ arising from the interaction of the particle with the homogeneous background. Using a compensating background for $p>3$ is not necessary, and as we show below, it is in fact not possible, as such a term would diverge.
    \item This additional interaction $V_{i,\mathrm{charge}}$ has been computed explicitly in the previous subsection, and can be seen to be finite for $p<3$ and divergent for $p\geq 3$. Thus, one approach is to use a compensating background for $p<3$ and to properly take into account the additional interaction with the background charge using $V_{i,\mathrm{charge}}$, while not to use any background at all for $p>3$, and hence also $V_{i,\mathrm{charge}}$.
    \item For the critical case $p=3$, charge neutrality does need to be enforced, but due to the divergence of the interaction between a potential compensating background, it is not possible to account for this. We thus set $V_{i,\mathrm{charge}}=0$, which essentially means that for this special case, we are not fully taking into account the interaction between the background and the particles.
    \item To be more consistent and ensure continuity of the potentials as the exponent $p$ is varied, it would be an option to also set $V_{i,\mathrm{charge}}=0$ for $p<3$. While this is a possibility, we decided against this as for the most important case of the Coulomb potential, it is common to include this term. We have therefore prioritized consistency with other Ewald implementations for $p=1$. It should be easy for power users to change this to any other desired convention.
\end{enumerate}

%% file: si/sections/lode.tex
\section{Relation to LODE and Evolution of Method}
\label{sec:lode}

We now discuss in some detail the relation between this work and some of our previous work for long-range ML, in which we have introduced the Long-Distance Equivariant (LODE) descriptors \cite{grisafi_incorporating_2019} and extensions thereof \cite{huguenin-dumittan_physics-inspired_2023}. While the framework presented in this work does also
support LODE, we recommend in most cases to use the simpler (exterior) potential
features. As we will now show, the majority of the long-ranged information of a system is contained in these features. For more complicated cases, however, there might be benefits to using the full LODE features.

The central tool that has driven these developments is a mathematical theorem called multipole expansion, which is discussed in detail in the supporting information of our previous work~\cite{huguenin-dumittan_physics-inspired_2023}.

The first half of this section discusses some of the key ideas that led from the original formulation~\cite{grisafi_incorporating_2019} to previous work published in 2023~\cite{huguenin-dumittan_physics-inspired_2023}. The second half then discusses the more recent developments that led to this work.

\subsection{First Two Mathematical Consequences: From Original to Previous Work}

We begin by discussing some of the inefficiencies in the original formulation of LODE
\cite{grisafi_incorporating_2019}, which have already been addressed in a previous publication
\cite{huguenin-dumittan_physics-inspired_2023}. We provide this background because our newer developments are
continuations of these improvements, and also to provide a more transparent
description of why we decided to seemingly change direction.

The LODE features (assuming a single chemical species for simplicity) are parametrized
by three indices $(n,\ell,m)$, $n=0,1,\dots,\nmax-1$, $\ell=0,1,\dots,\lmax$ and for each $\ell$,
$m$ takes the usual $(2\ell+1)$ distinct values $m\in\lbrace -\ell,-\ell+1,\dots,\ell\rbrace$.
Hence, for fixed $\lmax,\nmax$, the total number of LODE features is given by
\begin{align}
	\nmax \cdot (\lmax + 1)^2.
\end{align}
This number is directly proportional to the computational cost that is required to generate the features -- although when using a reciprocal-space implementation there is a large constant overhead associated with the global Fourier transform operations that might be dominant over the cost of computing individual features. 
If there are $N_a$ distinct chemical species in the system, this number is further multiplied by $N_a$. Hence, to reduce the computational cost of the
already expensive long-range part, it is essential to make sure that the relevant
long-range information is concentrated in as few coefficients as possible. It was shown previously that
the original method proposed in \cite{grisafi_incorporating_2019} is not doing this efficiently.

\begin{SCfigure}[][b]
	\label{fig:feature_visualization_all}
	\includegraphics[width=0.5\textwidth]{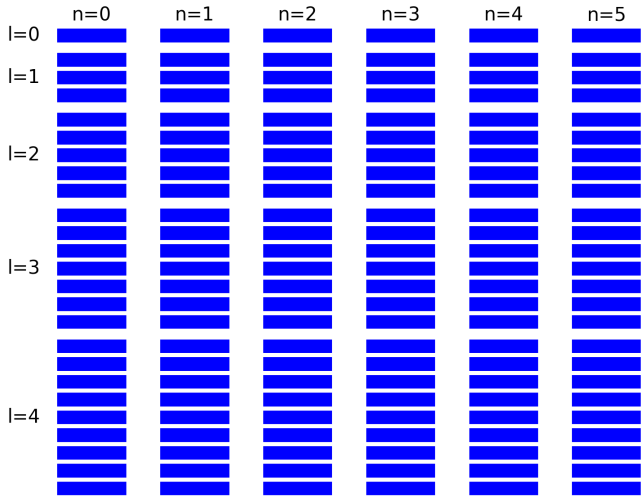}
	\caption{Visualization of features for $\nmax=6$ and $\lmax=4$ for the Coulomb potential. Each of the $\nmax
	\cdot (\lmax + 1)^2=150$ blue blocks corresponds to one feature that is computed
	when working with the LODE descriptor. Thus, the computational cost to get the LODE
	features is directly proportional to this number.\vspace{14em}}
\end{SCfigure}

To guide the following discussion, we will use $\nmax=6$ and $\lmax=4$ as an example,
for which we obtain $\nmax \cdot (\lmax + 1)^2=150$ features. A visual representation of
all those features is presented in \autoref{fig:feature_visualization_all}. Each of the
$150$ blue boxes corresponds to one feature, i.e. one set of possible values $(n,\ell,m)$, ordered according to the radial index $n$ in the horizontal and angular index $\ell$ in the vertical direction.
The number of boxes is proportional to the computational cost of evaluating the
features, and we will now examine how much long-range information is actually contained
in these.

\subsubsection{First Consequence: Most Features Contain No Long-Range Information}

We begin our discussion with the Coulomb potential. The key mathematical tool for this
entire discussion will be the multipole expansion theorem, which is discussed in detail
in the supporting information of \cite{huguenin-dumittan_physics-inspired_2023}. The discussion here will however be
much more high-level, meaning that it is not necessary to understand the mathematical
details of the theorem. It is important to note, however, that the results have been derived using open boundary conditions (for a charge density present in the entire space $\real^3$ decaying sufficiently quickly to be more precise), and not for periodic boundary conditions.

The first consequence of the multipole expansion theorem is that the entire long-range
information can be condensed into a single radial channel, e.g. into the coefficients
where $n=0$.

\begin{SCfigure}
	\label{fig:feature_visualization_noLR}
	\includegraphics[width=0.5\textwidth]{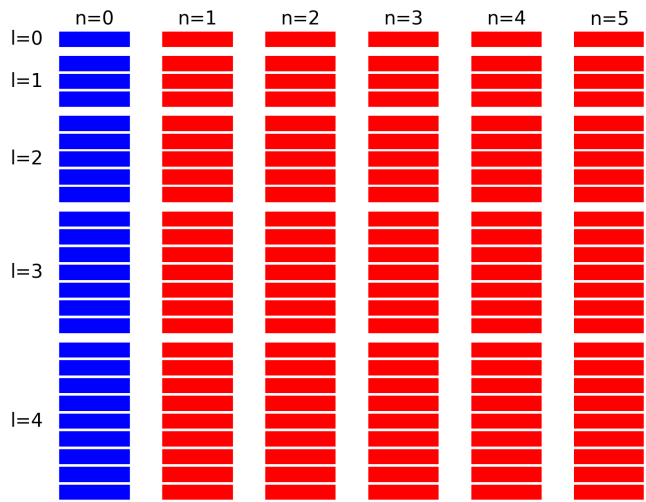}
	\caption{Visualization of LODE features for $\nmax=6$ and $\lmax=4$ for the Coulomb potential. The color corresponds to the amount of long-range information in the coefficient, with red meaning exactly zero information. This visualization shows that most features apart from the column with $n=0$ are completely useless: they do not contain any long-range information.\vspace{14em}}
\end{SCfigure}

This is visually represented in \autoref{fig:feature_visualization_noLR} for our working
example: out of the 150 coefficients that we needed to compute, it turns out that only
$25$ contain long-range information, while the remaining 125 or $\frac{5}{6}\approx
83\%$ in fact contain no long-range information at all! What this means in more
practical terms is that if we keep the position of all atoms within a cutoff, but move
the atoms outside of the cutoff, all features for $n>0$ will be invariant under this
operation. The coefficients for $n>0$ therefore only depend on the position of the interior atoms.

Since the computational cost to evaluate the LODE features (after performing the reciprocal-space part of the algorithm) is proportional to the number
of coefficients, we are essentially wasting $\frac{5}{6}\approx 83\%$ of computational
cost to compute some numbers that are pure short-range descriptors, the only difference
being that the implementation is using a much more sophisticated and expensive algorithm. 
Given that long-range features are usually combined with another implementation that provides
short-range descriptors, we are essentially just wasting computational effort. For a general value of $\nmax$, the fraction of wasted computational effort is $1-\frac{1}{\nmax}$. Note that this discussion only applies to the Coulomb potential, while the
more general case of a power-law potential is discussed later.

This provides us with a simple and immediate optimization: we can simply stop computing
features for $n>0$, and only focus on $n=0$. This leads to an improvement of the
computational cost by a factor of typical values of $\nmax$, which is on the order of
$6\sim 12$ for many applications, and hence about an order of magnitude.

\subsubsection{Second Consequence: Short-Range Contamination}

\begin{SCfigure}
	\label{fig:feature_visualization_contamination}
	\includegraphics[width=0.5\textwidth]{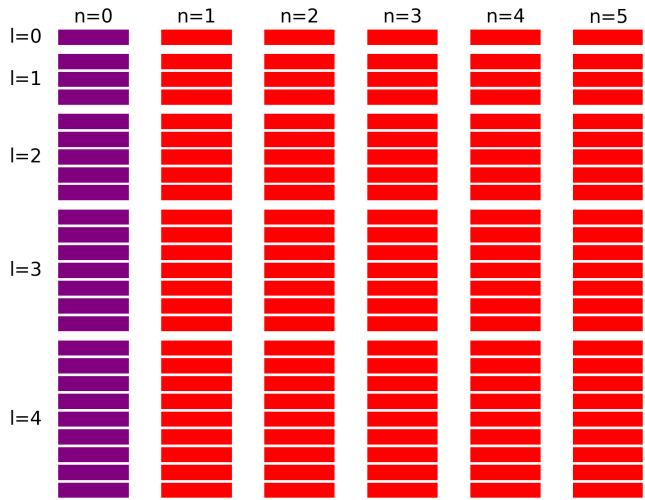}
	\caption{Visualization of features for $\nmax=6$ and $\lmax=4$ for the Coulomb potential. The color corresponds to the amount of long-range information in the coefficient, with red meaning exactly zero information, and purple meaning that the information is mostly-short ranged, with a tiny hint of long-range. We can therefore see that out of the $150$ coefficients, essentialyl all of them are completely useless for any realistic task in long-range ML.\vspace{12em}}
\end{SCfigure}

The second key deficit of the original LODE formulation, as well as many alternative approaches to long-range ML as we discussed in the main text, is the fact that even the slowly-decaying Coulomb potential is primarily determined by short-range contributions from immediate neighbors (see also fig. 2 in the main text). While this is fine if our goal is to learn an actual Coulomb potential, this makes the methods less suitable for more general long-range ML.

What this means for our working example of LODE with $\nmax=6$ and $\lmax=4$ is visually illustrated in \ref{fig:feature_visualization_contamination}. We had already identified the column for $n=0$ as the only one
containing long-range information, but even this is mostly just short-range information
with a tiny bit of long-range on top. This means that in the original formulation (and
implementation) of LODE \cite{grisafi_incorporating_2019}, a user would have spent a large
computational cost to evaluate 150 features, 125 of which contain exactly zero
long-range information, while the remaining 25 are also mostly short-range information.

While a sufficiently complex neural network might in principle be able to extract some long-range contributions from the 25 features, this signal is hidden behind the much larger short-range contributions that act as noise, making it unrealistic for this to happen in realistic datasets.

As we also recommend in the main text, short-ranged models today are quite mature, and
much more efficient at capturing short-range information. Furthermore, it is also this
short-range part that contains the dominant part of interactions, with long-range parts
being corrections. This is why we always recommend to use long-range features in
combination with a short-range ML part. In particular, this means that the $n=0$ components
of the LODE features, which are mostly short-ranged with a tiny fraction of long-range
information, are unlikely to improve model performance by any significant extent.

The only exceptions to this general guideline are toy systems that have been constructed
specifically to be challenging for short-range models. The original papers mostly
discussed such systems, but did not include more realistic target materials, which made
it harder to realize that LODE was essentially just computing short-range features in a
very expensive manner.

The solution to this problem is the subtraction of the interior contributions. This
makes the potential ``unphysical'', but leads to descriptors that only contain
long-range information. This is analogous to the ``purified'' long-range descriptors, also called exterior potential features (EPFs) in this work.

Both of these issues have already been addressed in our previous work \cite{huguenin-dumittan_physics-inspired_2023}, and the suggested solutions been implemented.

\subsection{Further Mathematical Consequences: From Last Publication to Current One}
The two mathematical consequences and possible optimizations obtained from the multipole expansion theorem discussed so far had already been addressed in our last publication \cite{huguenin-dumittan_physics-inspired_2023}. We now discuss further mathematical consequences from the theorem, which strongly influenced the direction taken in this work.

\subsubsection{Third Consequence: Not all coefficients are equal}
So far, we have argued that only the LODE coefficients for $n=0$ contain long-range
information (consequence 1) and that even extracting this information requires a purification process (consequence 2, also discussed in the main text). At the end of these steps, we still have to compute for each pair $(\ell,m)$, where $\ell=0,1,\dots,\lmax$ and $m\in\lbrace -\ell,\dots,\ell\rbrace$. The third consequence from the multipole expansion theorem is the fact that the
``amount'' of long-range information in each of those coefficients depends
on $\ell$. Note that the discussion here is still restricted to the Coulomb potential, with
the general case delegated to the last subsection.

Stated slightly more mathematically, the third consequence states that in a LODE coefficient corresponding to the indices $(\ell,m)$, the contribution of an atom at a distance $r$ is proportional to
\begin{align}
    \frac{1}{r^{1+\ell}}\,.
\end{align}
i.e. decaying faster for larger indices $\ell$. In particular, this reduces to $1/r$ for $\ell=0$, which is not surprising given that we are working with
the Coulomb potential to begin with. For $\ell=1$, the contribution of an atom
decays as $1/r^2$. This is the same decay as the electric field, and we will shortly
explain that this is not a coincidence (the three coefficients for $\ell=1$ basically are the components of the electric field). More generally, as we keep making $\ell$ large, the
atomic contributions start to decay faster and faster. In particular, using a large
angular index of $\ell=8$ would lead to contributions decaying as $1/r^9$, which can hardly
be characterized as long-range.

\begin{SCfigure}
	\label{fig:feature_visualization_lr_info}
	\includegraphics[width=0.5\textwidth]{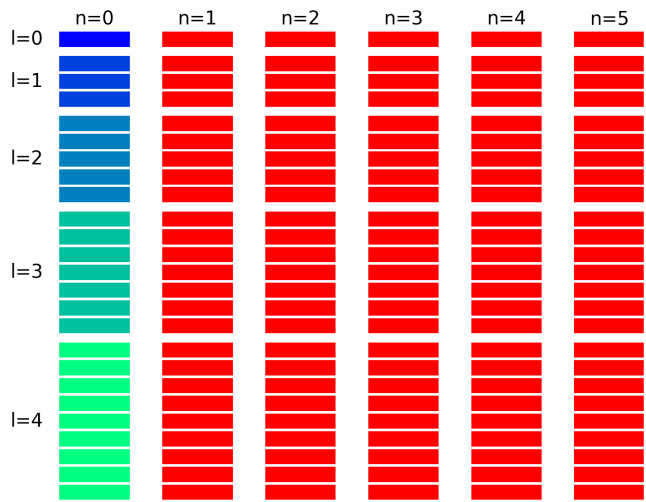}
	\caption{Visualization of LODE features and $\nmax=6$ and $\lmax=4$. The color of the block shows the amount of long-range information in the coefficient: the bluer the color, the more information there is. We can see that most long-range information is contained in $n=0$ and $l=0,1,2$.\vspace{15em}}
\end{SCfigure}

\autoref{fig:feature_visualization_lr_info} contains a visual representation of this
fact. The bluer the color, the more long-ranged the information in the
coefficients. This tells us that if the goal is to have a multi-scale model that
combines short- and long-range features, it will suffice to only use LODE coefficients
of small $\ell$, e.g. only $\ell=0,1,2$.

\subsubsection{Fourth Consequence: Relations between LODE Features, Potentials and Electric Fields}
The fourth consequence from the multipole expansion theorem is that in a certain limit (namely the one in which the cutoff radius $r_\mathrm{cut}\rightarrow 0$), the LODE coefficient for $\ell=0$ (and hence $m=0$) becomes equal to the electrostatic potential at the position of the center atom, while the three LODE coefficients for $\ell=1$ (and hence $m=\lbrace -1,0,1\rbrace$) become equal to the three components of the electric field at the position of the center atom. Note that to be more precise, both ``equalities'' are only true up to a global prefactor which can be computed analytically, and depends on the choice of LODE basis functions.

This fourth consequence is consistent with our previous observation, that the contribution of an atom at a distance $r$ decays as $1/r^{1+\ell}$, as this leads to $1/r$ for $\ell=0$ and $1/r^2$ for $\ell=1$. Note that the electric field is just the first derivative (or gradient to be more precise) of the potential. Focusing just on the radial dependence, $1/r^2$ is the first derivative of $1/r$ with respect to $r$ up to a global prefactor of $(-1)$.

More generally, the $\ell$-th derivative of $1/r$ is $1/r^{1+\ell}$ up to a prefactor, which is precisely the contribution of an atom to a LODE coefficient with index $\ell$. We might hence guess that more generally, the LODE components for some $\ell$ are related to the $\ell$-th derivatives of the potential. This guess turns out to be correct, and can be proven using the connection between the spherical and Cartesian versions of the multipole
expansion.

Thus, if implemented using an
autodifferentiable framework such as \torch or \jax, one can in principle obtain all
LODE coefficients just from the EPFs via suitable differentiation.

In other words, it suffices to implement the EPFs (the $\ell=0$ part), which will automatically provide us with the components for $\ell=1$ by taking a single differentiation step, $\ell=2$ by using second derivatives etc. This procedure clearly becomes less and less efficient for larger $\ell$, but shows that in principle,  our framework is fully able to generate the LODE features without too much extra effort. The main text also discusses some additional capabilities that have been added to also compute the features for higher angular indices $\ell$ efficiently.

Despite of this possibility, we do not recommend to use higher-$\ell$ coefficients too extensively. This is due to the previous observation that the larger $\ell$ becomes, the faster the contributions of the far-away atoms decay. This is why in this work, we suggest alternative methods to extract long-range information more efficiently: by modifying the particle weights (or charges) $q_i$ and the interaction exponent $p$. The next subsection discusses how some of the results discussed so far change when moving to a more general exponent $p$.

\subsection{Mathematical Results for General Exponents $p$}
The results so far have exclusively been about the Coulomb potential, which is the most important long-range interaction. Both the LODE features and the EPFs introduced in this work can however easily be generalized to arbitrary $1/r^p$ inverse power-law potentials with $p>0$ being a positive real number.

As discussed extensively in the supporting information of our previous work \cite{huguenin-dumittan_physics-inspired_2023}, the multipole expansion theorem for this general case is quite different from the Coulomb potential, which is the special case for $p=1$. It might at first seem surprising that among all positive reals, the Coulomb potential only is very different. We thus explain this first in the next subsection. Readers only interested in the consequences for ML models can skip directly to the final part.

\subsubsection{Generalized Multipole Expansion}
 The special nature of the Coulomb potential $v(r) = 1/r$ arises from the fact that it is a solution to the Laplace equation
\begin{align}
    \Delta v = \nabla^2 v = 0
\end{align}
for $r \neq 0$, where $\Delta = \nabla^2$ is the Laplace-operator. In fact, it is the unique spherically symmetric solution to this equation apart from the constant function! This defining property of the Coulomb potential, perhaps surprisingly, has a dramatic impact on its multipole expansion.

Remembering the definition of the Laplace operator, this means that a certain linear combination of second derivatives of $v$ vanishes. If we perform a Taylor expansion (in three dimensions) of a generic smooth function $v$, the quadratic term consists of $6$ independent terms proportional to $x^2, y^2, z^2, xy, xz, yz$ respectively (since $xy=yx$ commute, as do the derivatives). The prefactors of these terms are precisely the six second derivatives of the target function with respect to two of the three variables. If a function $v$ is a solution to the Laplace equation, however, this means that a linear combination of these six second derivatives is zero, and hence that only five of the derivatives are actually independent. This is why for the Coulomb potential, the contributions with $\ell=2$ decaying as $1/r^{1+\ell}=1/r^3$ only contain five terms: $m\in\lbrace -2,-1,0,1,2\rbrace$. The pattern of vanishing derivatives propagates to higher-order derivatives, which means that for the $\ell$-th derivatives, only $2\ell+1$ are independent. This is what makes the Coulomb potential so special when it comes to its Taylor expansion.

For a more generic function, i.e. a function that is not a solution of the Laplace equation, there are no such cancellations, and in general at order $\ell$, there ar%
\begin{align}
    \begin{pmatrix}
        \ell + 2 \\ 2
    \end{pmatrix} = \frac{(\ell+2)(\ell+1)}{2} \sim \ell^2
\end{align}
independent terms in its Taylor expansion. This Taylor expansion is also called ``Cartesian multipole expansion'', as all the terms of $\ell$-th derivatives decay as $1/r^{1+\ell}$ (with some angular dependence), just as for the usual or ``spherical'' multipole expansion.

So far, we have explained how the Taylor or Cartesian multipole expansion of a generic $1/r^p$ potential contains a lot more terms than $1/r$. The final step to fully make the connection to the LODE coefficients is to connect this Cartesian multipole expansion to the usual ``spherical'' multipole expansion. This last step simply requires the realization that a spherical harmonic function $Y_\ell^m$ is, up to a factor of $r^\ell$, just a polynomial of degree $\ell$ in the Cartesian coordinates $(x,y,z)$. Hence, going from the Cartesian coordinates to spherical harmonics is simply a linear change of basis, which shows that both the Cartesian and spherical multipole expansions are equivalent term-by-term up to a change of basis. The more explicit statements are discussed in the next subsection.

\subsubsection{Fifth consequence: Results are More Complicated But Similar for General Inverse Power-Law Potential}

For LODE features generated from a general $1/r^p$ inverse power-law potential with $p\neq 1$, the information content in the LODE coefficients behaves differently from the Coulomb potential. The generalized multipole expansion discussed in the supporting information in our previous work \cite{huguenin-dumittan_physics-inspired_2023} shows that it is not possible to condense the entire long-range information into $n=0$, and that all values of $n$ are relevant. Hence, there are now relevant LODE coefficients for all values of the indices $(\ell,m,n)$ rather than just $(\ell,m)$.

\begin{SCfigure}
	\label{fig:feature_visualization_general_ipl}
	\includegraphics[width=0.5\textwidth]{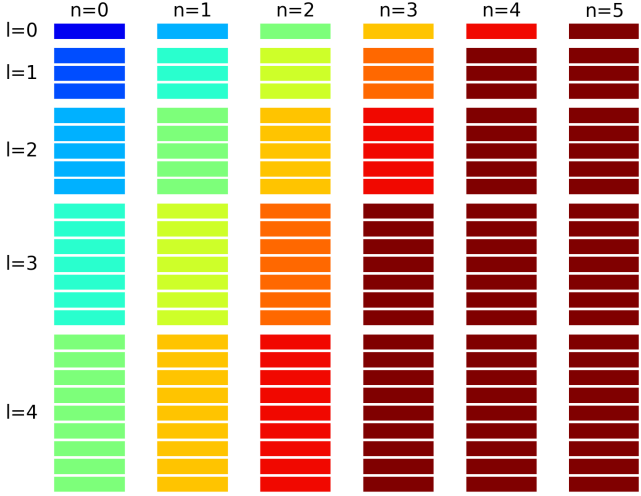}
	\caption{Visualization of LODE features and their long-range information content for $\nmax=6$ and $\lmax=4$ for a general $1/r^p$ inverse power-law potential. The color of the block corresponds to the amount of long-range information in the coefficient. We see that the long-range information decreases for large $n$ and $\ell$.\vspace{14em}}
\end{SCfigure}

For a given coefficient, the contribution of an atom at a distance of $r$ from the center then decays as
\begin{align}
    \frac{1}{r^{p+\ell+2n}}\,.
\end{align}
This shows more directly that all radial channels $n$ are important, but also that the decay is much faster for large $n$. Hence, the general trends are still similar to the Coulomb potential: The majority of the long-range information is contained in the coefficients with $n=0$ and small values of $\ell$ such as $\ell=0,1,2$. The only difference is that now, also the coefficients with $n=1,2$ contain some long-range information. This is visualized in \autoref{fig:feature_visualization_general_ipl}.

The relation between the potential (or EPFs) and the analogue of the electric field are the same as for the Coulomb potential. The LODE coefficient for $n=0$ and $\ell=0$ still corresponds to the potential (or EPF) at the location of the center atom. Furthermore, the three ``generalized electric field'' coefficients with $n=0$ and $\ell=1$ (and hence $m\in\lbrace -1,0,1\rbrace$) by taking the gradient of the coefficient with $n=0$ and $\ell=0$.

Despite of this more complicated nature, these facts and the fast decay with respect to $n$ means that the conclusions are mostly similar to the Coulomb potential. This is why instead of building expensive LODE features by using multiple values of $\ell$ and $n$, we suggest to just use the potential itself (which essentially corresponds to LODE with $n=\ell=m=0$) with different charges and the option to add gradients (essentially adding the coefficients for $\ell=1$) either using the full or exterior variant.

%% file: si/sections/accuracy.tex
\section{Tuning the Ewald and PME parameters}
\label{sec:accuracy}

In the following, we discuss the procedure we implement to tune the optimal parameters, by minimizing
the absolute force error estimation. We define the total force error as
\begin{align}
  \Delta F &= \sqrt{\Delta F_\mathrm{SR}^2 + \Delta F_\mathrm{LR}^2}\,
\end{align}
where $\Delta F_\mathrm{SR}$ and $\Delta F_\mathrm{LR}$ are the absolute real-space and
Fourier-domain errors. 
Giving expressions for the expected error of general structures is
difficult and usual error estimations assume homogenous random systems.

For all methods the real space error is\cite{Kolafa1992Jan}
\begin{align}
  \Delta F_\mathrm{SR}
  \approx \frac{Q^2}{\sqrt{N}}
  \frac{2}{\sqrt{r_{\text{cut}} V}}
  e^{-r_{\text{cut}}^2 / 2 \sigma^2}
\end{align}
where $N$ is the number of charges, $Q^2 = \sum_{i = 1}^N q_i^2$, is the sum of squared
charges, and $V$ is the volume of the system.

For Ewald, the approximate force error is\cite{Kolafa1992Jan, arnold_methods_nodate}
\begin{align}
  \Delta F_\mathrm{LR}^\mathrm{Ewald}
  \approx \frac{Q^2}{\sqrt{N}}
  \frac{\sqrt{2} / \sigma}{\pi\sqrt{2 V / h}} e^{-2\pi^2 \sigma^2 / h ^ 2}
\end{align}

For PME it reads \cite{petersenAccuracyEfficiencyParticle1995}
\begin{align}
  \Delta F_\mathrm{LR}^\mathrm{PME}
  \approx 2\pi^{1/4}\sqrt{\frac{3\sqrt{2} / \sigma}{N(2P+3)}}
  \frac{Q^2}{L^2}\frac{(\sqrt{2}H/\sigma)^{P+1}}{(P+1)!} \times
  \exp{\frac{(P+1)[\log{(P+1)} - \log 2 - 1]}{2}} \left< \phi_P^2 \right> ^{1/2}
\end{align}
where $P$ is the order of the interpolation scheme, $H$ is the spacing of mesh
points, and $\phi_P^2 = H^{-(P+1)}\prod_{s\in S_H^{(P)}}(x - s)$, in which $S_H^{(P)}$ is
the $P+1$ mesh points closest to the point $x$.

For P3M the error is\cite{desernoHowMeshEwald1998a}
\begin{align}
  \Delta F_\mathrm{LR}^\mathrm{P3M}
  &\approx \frac{Q^2}{L^2}\left(\frac{\sqrt{2}H}{\sigma}\right)^P
  \sqrt{\frac{\sqrt{2}L}{N\sigma}
  \sqrt{2\pi}\sum_{m=0}^{P-1}a_m^{(P)}\left(\frac{\sqrt{2}H}{\sigma}\right)^{2m}}
\end{align}
where $a_m^{(P)}$ is an expansion coefficient, and its exact value can be found in Table
II in Ref.~\citenum{desernoHowMeshEwald1998a}.

Based on the error formulas above, and for a given trial structure that is representative of those most commonly used for the problem at hand, we can tune the parameters to achieve the desired 
accuracy following the algorithm below.

\begin{itemize}
  \item[Step 1] Based on the user given accuracy $\epsilon_\mathrm{target}$ and $r_\mathrm{cut}$, calculate the smearing $\sigma$ by solving $\Delta F_{\mathrm{SR}}(\sigma ; r_\mathrm{cut}) = \epsilon_\mathrm{target}/2$.
  \item[Step 2] Initialize sets of $k$-space parameters. For Ewald, the values are chosen such that each spatial direction (x, y, z) contains between 1 and 13 mesh points. For mesh-based methods (P3M and PME), the grid sizes range from 2 to 7 mesh points per spatial direction, considering all possible interpolation orders.
  \item[Step 3] Evaluate all possible parameter sets. The error associated with the parameter is estimated cheaply using the error formulas. Parameters that yield error within the desired accuracy are benchmarked estimating the empirical timings for an evaluation of energy and derivatives. Parameter sets exceeding the error threshold are not tested for performance and their timings are set as infinity.
  \item[Step 4] Return the parameter set achieving the desired accuracy with the shortest timing. If no parameter set can fulfill the accuracy requirement, the set achieving the smallest error is returned.
\end{itemize}

\begin{SCfigure}
  \includegraphics{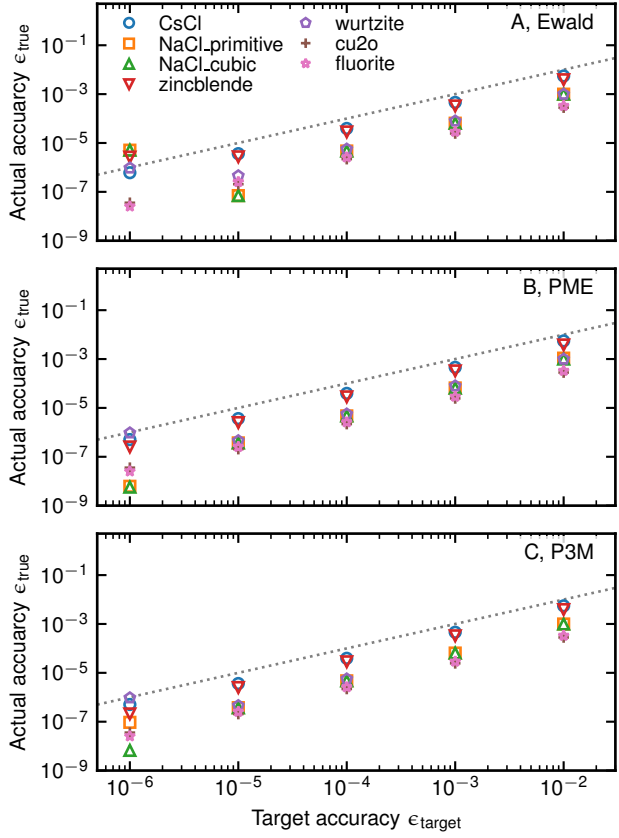}
  \caption{\label{fig:tuning}
    Similar as Figure 3 in the main text, this figure is showing the accuracy obtained after performing the tuning procedure for a given target accuracy for different crystal structures, using double
    precision arithmetics. Panel A shows the results for Ewald, panel B for PME module and panel C
  for P3M. \vspace{28em}}
\end{SCfigure}

In \autoref{fig:tuning} we show the results of the tuning procedure for Ewald, PME, and
P3M similarly to Figure 3 in the main text but for double precision (64-bit) floating point numbers. As
already described we find also that we achieve the requested accuracy for the Madelung
constants but the accuracy saturates around $10^{-4}$ for Ewald summation and $10^{-5}$
for the mesh calculators.

\begin{SCfigure}
  \includegraphics{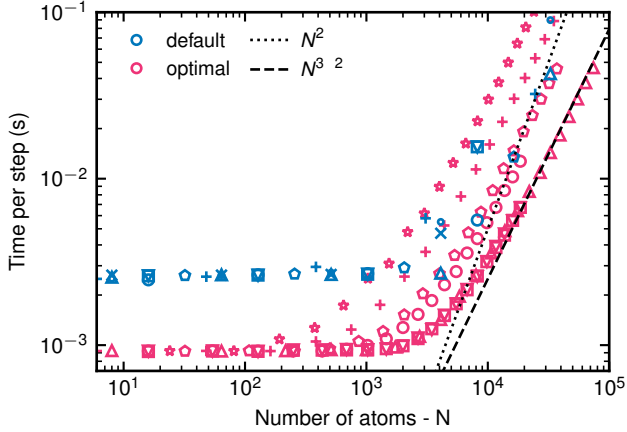}
  \caption{\label{fig:optimal_ewald} Benchmark for the computational costs of Ewald
    calculators. Blue symsbols show quadratic scaling as the Ewald parameters where
    optimized using the procedures given above. Red symbols show $N^{3/2}$ scaling using the
  fixed parameters given in \autoref{eq:ewald_optimal}. \vspace{12em}}
\end{SCfigure}

In figure \autoref{fig:optimal_ewald} we show the optimal parameters for Ewald summation
which leads to a scaling proportional to $N^{3/2}$ together with the quadratic scaling
as it is shown in the main text in Figure 4A. The optimal scaling
was achieved by setting
\begin{align}\label{eq:ewald_optimal}
  \mathrm{smearing} &= 1.3 \cdot N^{1 / 6} / \sqrt{2} \nonumber \\
  \mathrm{lr\_wavelength} &= 2 \pi \cdot \mathrm{smearing} / 2.2 \\
  \mathrm{r_c} &= 2.2 \cdot \mathrm{smearing}\nonumber
\end{align}
The parameters and pre-factors were obtained from systematic tests on a $\mathrm{CsCl}$ system, where the unit cell is replicated 16 times along each spatial dimension, yielding a total system of 8192 atoms.

%% file: si/sections/cutoff_search.tex
\section{Finding the optimal cutoff for benchmarking}

\begin{SCfigure}[][h]
  \includegraphics{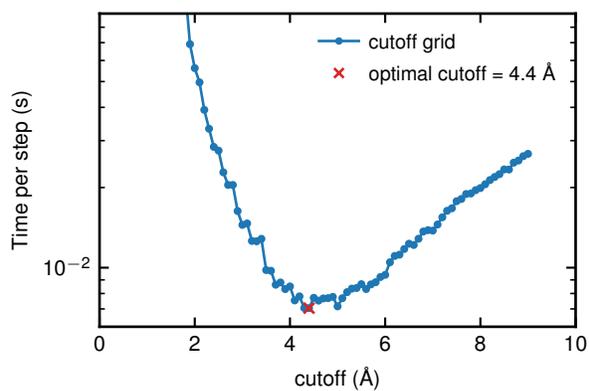}
  \caption{\label{fig:grid-search}Time per step vs real space cutoff for replicated NaCl
    structure in a cubic cell. The red cross indicates the best performing cutoff of
    $4.4$\,Å.\vspace{13em}}
\end{SCfigure}

To find the optimal cutoff for our benchmark simulations we perform a grid search over
the real space cutoff for a replicated NaCl structure in a cubic cell. We replicated the
initial cell 32 in each direction leading to a total of 262144 atoms in a cell with a
side length of 64\,Å. We use the same LAMMPS P3M setup as in the main text, but we only
vary the real-space cutoff. Figure \ref{fig:grid-search} shows the time per step versus
the real space cutoff, showing the optimum efficiency is achieved for  a value of $4.4$\,Å.

%% file: si/sections/gpu-utilization.tex
\section{GPU Utilization}
\label{sec:gpu_util}

\begin{figure}[h]
    \includegraphics{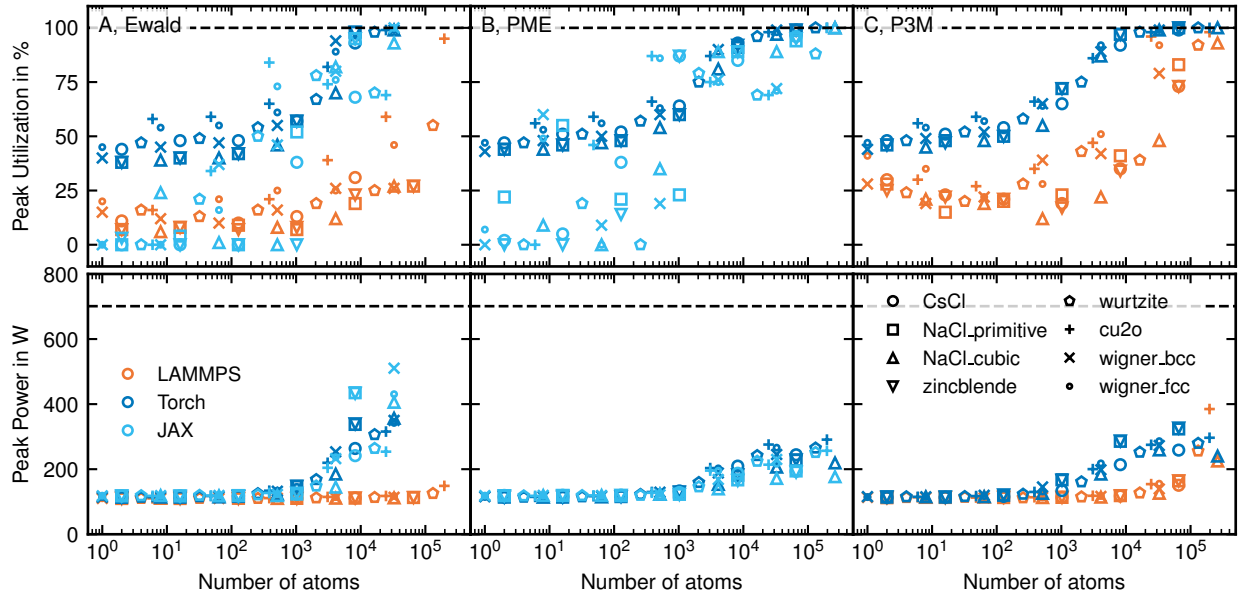}
    \caption{\label{fig:gpu-utilization}
    Upper panels: Peak percent utilization for Ewald (A), PME (B) and P3M (C) on a single NVIDIA H100 GPU. Lower panels: Peak power in watts for the same methods. Different crystals are shown as different symbols. Each point shows the maximal utilization or power during a run of 50 calculations of the same positions with a tuning accuracy of $1 \cdot10^{-4}$.}
\end{figure}

\rev{
During the benchmarks discussed in section IV A of the main text, we collected statistics on GPU utilization (the percentage of time during which at least one kernel is executed) and power draw (power consumed by the GPU) to estimate the extent to which our code, as well as the LAMMPS reference implementation, can make use of the available hardware. The results are shown in \autoref{fig:gpu-utilization}.

At system sizes below approximately $10 000$ atoms, the GPU is not fully utilized. Even at large sizes, neither our code nor LAMMPS is able to fully exhaust the available hardware, as shown by the power draw not reaching the peak design capacity of $700$~W. Therefore, we expect that significant further performance improvements can be achieved by more specifically targetting specific GPUs, for instance by overlapping memory access and compute, reducing overhead, and making use of reduced-precision operations where appropriate.
}

%% file: si/sections/water-nve.tex
\section{MD of Water at constant energy and volume}
\label{sec:water-nve}

\begin{figure}[h]
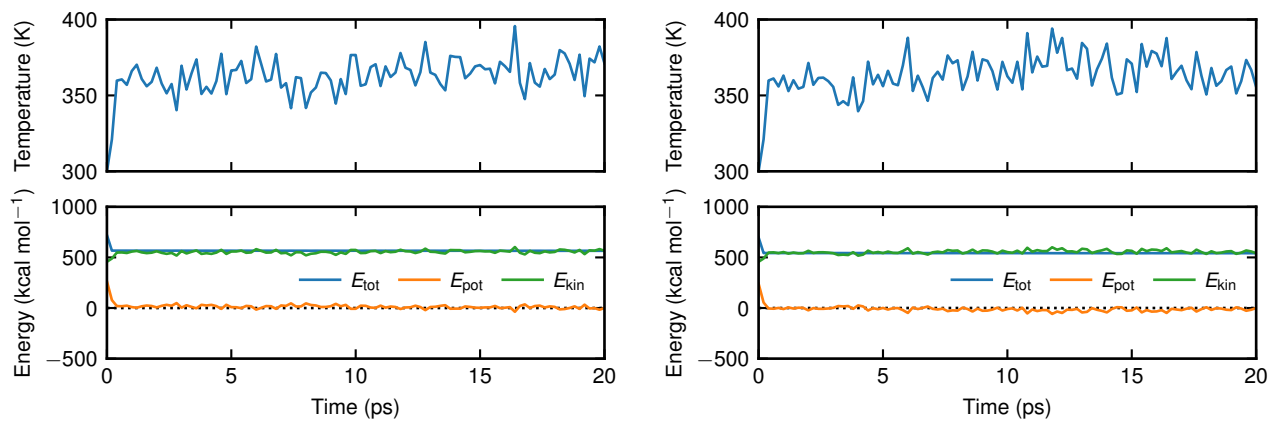

  \includegraphics{si/figures/water-nve-single}
  \includegraphics{si/figures/water-nve-double}
  \caption{\label{fig:water-nve}
  The temperature and the energy (split into kinetic and potential energy) of the water
  system in the NVE ensemble over $20\,\mathrm{ps}$. The left panels show the results
  for the Coulomb energy computed in single precision (32-bit floating-point precision),
  while the right panels show the results for the Coulomb energy computed in double
  precision.}
\end{figure}

Similar to the main text shown in Figure 5, we perform MD simulations of
the water system without thermostats at constant energy and volume (NVE) over
$20\,\mathrm{ps}$ using the same settings as described in section IV B of
the main text. Figure \ref{fig:water-nve} shows the temperature and the energy of the
system. We perform two simulations: in one, we compute the Coulomb energy in single
precision (32-bit floating-point numbers), and in the other, we use double precision. In
both cases, the energy is conserved, verifying that the forces are the exact derivative
of the energy up to numerical precision.

